%% file: iclr2027_conference.tex
\documentclass{article} 
\usepackage{iclr2027_conference,times}

\input{math_commands.tex}

\usepackage{hyperref}
\usepackage{url}
\usepackage{graphicx} 
\usepackage[utf8]{inputenc}
\usepackage{multirow}
\usepackage{pifont}
\usepackage{enumitem}

\newcommand{\cmark}{\ding{51}} 
\newcommand{\xmark}{\ding{55}} 

\usepackage{xcolor}
\newcommand{\swap}[1]{\textcolor{black}{#1}}
\newenvironment{swapblock}{\color{black}}{}
\definecolor{itadblue}{HTML}{2A78D6}
\definecolor{geminiorange}{HTML}{EB6834}
\definecolor{qwenaqua}{HTML}{1BAF7A}
  
\definecolor{Mycolor2}{HTML}{ccffff}
\definecolor{Mycolor3}{HTML}{dbdfde}

\usepackage{booktabs}
\usepackage{tcolorbox}

\newcommand{\modelname}{\texttt{TAC2}}
\newcommand{\studentname}{\texttt{TAC2-S}}
\newcommand{\teachername}{\texttt{TAC2-E}}

\title{\swap{TAC2: Timestamped Audio Captioning \\ as Composable Functions}}

\author{
\begin{tabular}[t]{@{}p{0.3\textwidth}p{0.3\textwidth}p{0.3\textwidth}@{}}
 Swapnil Bhosale\thanks{Equal contribution. Correspondence: s.bhosale@surrey.ac.uk , seethara@adobe.com}\hspace{0.45em}\thanks{\swap{Work done during internship at Adobe Research.}}
 & Prem Seetharaman\textsuperscript{\rm *}
 & Ke Chen \\
 University of Surrey, UK
 & Adobe Research, USA
 & Adobe Research, USA \\[12pt]
 Haonan Chen
 & Justin Salamon
 & Zeyu Jin \\
 Adobe Research, USA
 & Adobe Research, USA
 & Adobe Research, USA
\end{tabular}
}

\iclrfinalcopy 
\begin{document}

\maketitle

\input{sec/0_abstract}

\input{sec/1_introduction}
\input{sec/2_related}
\input{sec/3_method}
\input{sec/4_experiment}

\input{sec/5_conclusion}
\input{sec/6_reprod}

\bibliography{iclr2027_conference}
\bibliographystyle{iclr2027_conference}

\appendix
\input{sec/appendix}

\end{document}

%% file: math_commands.tex
\usepackage{amsmath,amsfonts,bm}

\def\eqref#1{equation~\ref{#1}}

\def\1{\bm{1}}

\DeclareMathAlphabet{\mathsfit}{\encodingdefault}{\sfdefault}{m}{sl}
\SetMathAlphabet{\mathsfit}{bold}{\encodingdefault}{\sfdefault}{bx}{n}



%% file: sec/0_abstract.tex
\begin{abstract}
Real-world audio applications require precise, queryable representations grounded in physical time.
Yet modern Large Audio-Language Models (LALMs) increasingly favor conversational generation, 
producing free-form descriptions that conflate acoustic perception with higher-level reasoning and lack precise temporal grounding.
We introduce \swap{\modelname{}, a foundation model that casts timestamped audio captioning as composable functions to construct} granular, speaker-attributed acoustic timelines directly from in-the-wild audio.
To avoid prohibitively expensive manual annotation, 
we programmatically fuse complementary predictions from specialized weak teachers through time-registered IoU matching.
We model the resulting multi-granularity schema of timestamped events, speaker turns, and fine-grained speech attributes using an Instruction-Input-Output (IIO) completion format,
\swap{in which each task is a function of the audio, an instruction, and an optional input.
A single call decodes the full description autonomously, while composing calls lets earlier outputs or external priors, such as a reference transcript, condition later ones.}
We further extend \modelname{} with expressive speech attributes to obtain \studentname{}, 
using parameter-efficient specialists to elicit latent paralinguistic capabilities from the pre-trained backbone and distill them into a unified model.
We validate the resulting representation in two complementary settings.
Paired with frozen text-only LLMs, 
it yields strong spoken audio understanding across diverse reasoning backends, 
demonstrating the utility of decoupling fine-grained acoustic perception from downstream reasoning.
Used as paired supervision for text-to-audio generation, 
it supports semantically and temporally grounded reconstruction, 
together with explicit zero-shot control over event timing, speaker attributes, and word-level emphasis.
Project page: \url{https://swapb94.github.io/tac2-audio-describer/}
\end{abstract}

%% file: sec/1_introduction.tex
\section{Introduction}
\label{sec:introduction}
Human auditory perception separates overlapping acoustic signals before higher-level reasoning~\citep{bregman1990auditory}, grounding abstract semantics in physical perception~\citep{harnad1990symbol, bender-koller-2020-climbing}.
In contrast, modern Large Audio Language Models (LALMs) emphasize conversational generation~\citep{salmonn,qwen2audio,qwen3omni}, producing free-form text with limited structural grounding.
Yet applications such as 
semantic search, precise audio editing, and automated archiving 
require queryable, temporally grounded representations of events, speakers, and paralinguistic cues.

\begin{figure*}[h]
\centering
\includegraphics[width=\textwidth]{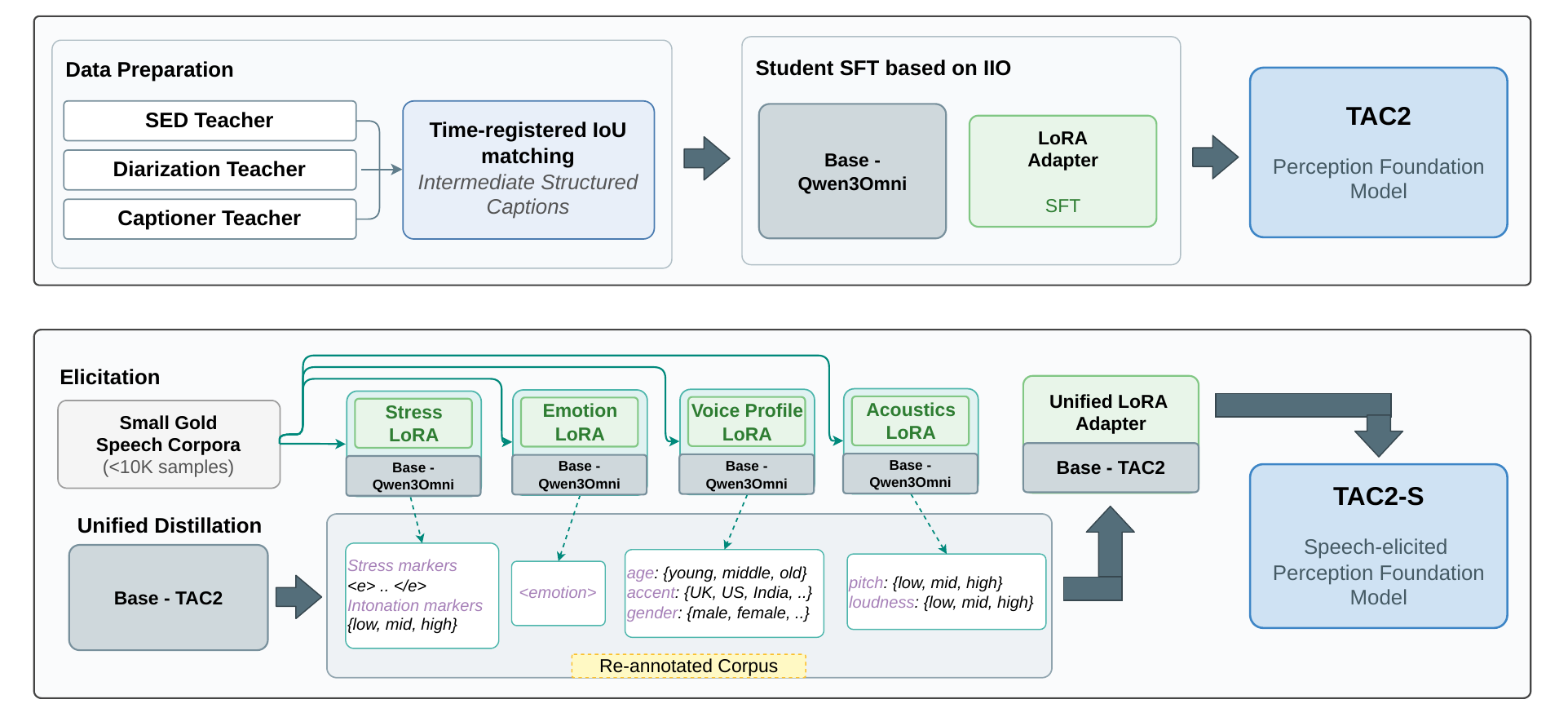}
\caption{\textbf{Overview of the two-stage \modelname{} / \studentname{} framework.} 
\textbf{(1) Perceptual Foundation (\modelname{}):} 
Predictions from weak teachers (SED, Diarization, Captioner) are merged via time-registered IoU matching to generate intermediate structured captions, 
which train a base adapter via the proposed Instruction-Input-Output (IIO) format.
\textbf{(2) Speech Elicitation (\studentname{}):}
Individual LoRA specialists
(Stress, Emotion, Voice Profile, Acoustics)
are trained on small gold corpora to elicit paralinguistic attributes from the pre-trained backbone.
This cascade re-annotates the corpus, which is subsequently distilled into our unified student model, \studentname{}.
}
\label{fig:overview}
\end{figure*}

Prior work on Timestamped Audio Captioning (TAC)~\citep{tac} took a first step toward this goal by grounding audio descriptions in physical time, 
representing acoustic events with explicit temporal boundaries. 
However, existing timestamped captioning formulates description generation largely as a fixed, one-shot process that does not naturally accommodate richer perceptual structure or external priors.
This motivates a natural question, 
\textit{Can richer, structured audio descriptions be \swap{built by composing simpler captioning functions, where the output of one conditions the next?}}

We propose \modelname{}, 
a foundation model that constructs 
granular, speaker-attributed acoustic timelines 
directly from in-the-wild audio.
To avoid manually annotating such structured timelines at scale,
we introduce a multi-view fusion pipeline that temporally aligns complementary predictions from off-the-shelf 
sound event detectors, diarizers, and captioners 
over a large real-world corpus, producing high-fidelity structural training targets.
To construct this richer schema while retaining modular control, we introduce an \textbf{Instruction-Input-Output (IIO)} completion format that decomposes the structured caption into conditionable completion tasks.
\swap{Each IIO task acts as a function of the audio, an instruction, and an optional input.
A single call decodes the full description, while composing calls lets preceding predictions or external priors, such as high-quality transcripts, condition subsequent completions.}
We further extend \modelname{} with expressive speech attributes to obtain \studentname{}, formulating this expansion as latent capability elicitation.
Motivated by probing evidence that these attributes are decodable from intermediate backbone representations, we train attention-only specialists on fewer than 10,000 labeled samples to expose them as structured outputs.
These specialists annotate the corpus with expressive attributes, and the resulting cascade is distilled into the unified \studentname{} model.

The resulting structured representation enables two downstream paradigms.
First,
cascading \studentname{} with frozen text-only LLMs yields strong spoken understanding on MMSU~\citep{mmsu} and MSU-Bench~\citep{msubench} across diverse reasoning backends,
bridging perception and reasoning through a structured intermediate representation.
Second,
using \studentname{} to annotate in-the-wild audio produces a large-scale, paralinguistically enriched structured training corpus for text-to-audio generation.
We show that this corpus enables explicit zero-shot control over event timing, speaker characteristics, and word-level emphasis.

In this work, our contributions are:
(1) a scalable supervision framework that constructs granular, temporally grounded training targets by fusing complementary weak teachers, and extends them with paralinguistic attributes elicited by lightweight specialists trained on small labeled datasets
(2) Instruction-Input-Output (IIO), a modular structured-completion interface that \swap{casts captioning tasks as composable functions, supporting both autonomous decoding and externally conditioned inference}, enabling flexible construction and bootstrapping of comprehensive audio descriptions; and
(3) \modelname{} and its paralinguistic extension \studentname{}, which instantiate this framework and demonstrate strong fine-grained audio understanding together with temporally and paralinguistically controllable audio generation.

%% file: sec/2_related.tex
\section{Related Work}
\label{sec:related_works}

\paragraph{Audio-Language Models and Grounded Perception.}
Recent Large Audio-Language Models (LALMs)~\citep{salmonn,qwen2audio,kong2024audio,zhang2025mimo} demonstrate strong instruction-following, 
yet benchmarks reveal a persistent gap between high-level reasoning and fine-grained acoustic perception~\citep{mmsu,mmau}. 
Existing models remain susceptible to temporal hallucinations and reliance on linguistic priors~\citep{kuan2025can,cheng2026aha},
a limitation compounded by captioning datasets that summarize audio into short global descriptions~\citep{audiocaps,clotho}.
While prior timestamped captioners such as TAC~\citep{tac} introduced explicit event boundaries, they primarily target event-level temporal grounding and are not designed to represent the richer structure of natural conversational audio, including speaker-attributed turns and overlapping events.
In contrast, \modelname{} uses multi-teacher temporal fusion to construct explicit, speaker-attributed acoustic timelines from in-the-wild audio, and represents them through IIO as a multi-granularity perceptual interface that supports both autonomous decoding and externally conditioned inference.

\paragraph{Spoken Paralinguistics and Latent Elicitation.}
Expressive speech attributes such as stress, intonation, emotion, and speaker characteristics are central to spoken understanding~\citep{expresso,speechcraft}. 
Existing SpeechLLMs often model these attributes as isolated categorical tasks~\citep{blsp,osum} or rely on external signal-processing pipelines for acoustic features~\citep{whistress,emotionthinker}. 
Meanwhile, work on latent capability elicitation suggests that pretrained models can encode capabilities that become accessible through limited parameter adaptation~\citep{aghajanyan2021intrinsic,donoway2025quantifying}. 
Building on this observation, \studentname{} uses lightweight specialists to expose paralinguistic information that is decodable from the pretrained audio backbone, and distills their predictions into a unified, multi-granularity acoustic schema.

%% file: sec/3_method.tex
\section{Methodology}
\label{sec:methodology}

Our framework constructs a queryable, temporally grounded acoustic representation in two stages.
Stage~1 builds \modelname{}, 
a granular, speaker-attributed physical timeline bootstrapped from complementary weak teachers over in-the-wild audio.
Stage~2 extends this timeline into \studentname{} by eliciting latent paralinguistic capabilities from the pre-trained backbone using lightweight specialists, then distilling their annotations into a unified model.

\subsection{Stage 1: Perceptual Foundation (\modelname{})}
\label{sec:stage1_method}
Dense temporal annotation at the scale required to train a foundation model is prohibitively expensive.
We therefore construct \modelname{}'s training targets through multi-view weak supervision and learn them using an Instruction-Input-Output (IIO) completion paradigm.

\begin{figure*}[h]
\centering
\includegraphics[width=\textwidth]{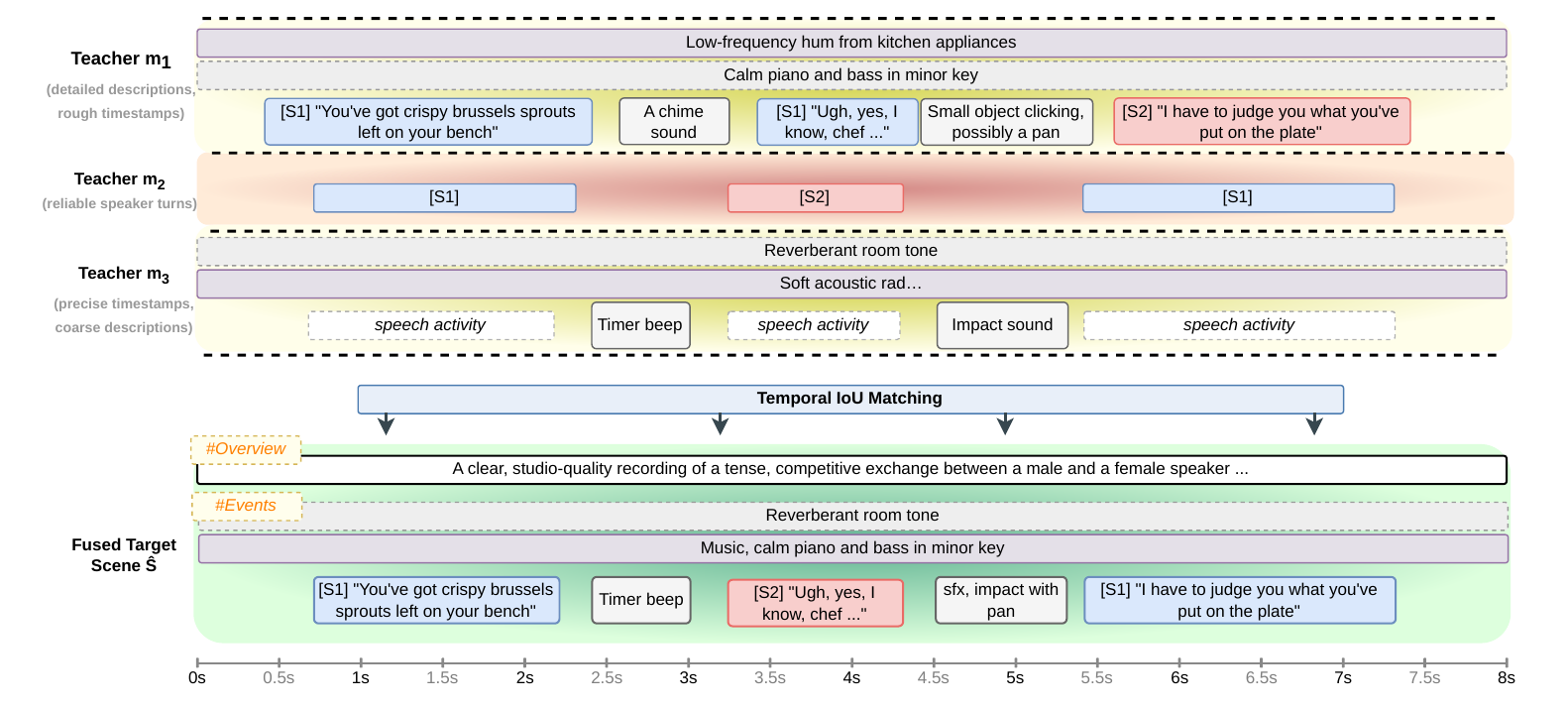}
\caption{
\textbf{Programmatic data compilation via time-registered IoU matching.}
Predictions from three independent weak teachers
($m_1$: audio captioner; $m_2$: speaker diarizer; $m_3$: sound event detector)
are aligned on a shared time axis using temporal IoU with overlap gate
(Eq.~\ref{eq:iou},\ref{eq:injective}).
Their aligned predictions form the fused target timeline $\hat{S}$ (Eq.~\ref{eq:fused_scene}), containing concurrent acoustic events with speaker attribution.}
\label{fig:iou_matching}
\end{figure*}

\paragraph{Multi-view Perceptual Fusion} 
We programmatically construct \modelname{}'s target timelines by fusing predictions from specialized weak teachers, 
including audio captioning, speaker diarization, and sound event detection.
While each teacher provides complementary task-specific predictions under a distinct schema, all predictions share physical time as a common coordinate.
We exploit this common axis to align their disparate temporal boundaries and attributes (Figure~\ref{fig:iou_matching}).

Let the latent scene $S$ comprise acoustic events over the attribute schema
$\mathcal{A} = \{\text{event type}, \text{acoustic description}, \text{temporal boundaries}, \text{speaker identity}, \text{transcript}\}$. 
Each of $K$ specialized teachers $m_k$ produces a partial view $O_k={o}$, with a temporal support $\tau(o)=[o_s,o_e]$.

To fuse these views, we define an authority map
$\text{auth} : \mathcal{A} \to \{1, \dots, K\}$ 
that maps each attribute to its most reliable teacher. 
Furthermore, each acoustic event class $c$ is assigned a temporal owner $\kappa(c)$ that dictates its master temporal segmentation. 
Temporal agreement between two predictions is measured using their temporal Intersection over Union (IoU):
\begin{equation}
\text{IoU}(o, o') = \frac{|\tau(o) \cap \tau(o')|}{|\tau(o) \cup \tau(o')|}
\label{eq:iou}
\end{equation}
For each event class $c$, let
$E_c = (e_1, \dots, e_{n_c})$
be its reference events drawn from its temporal owner $\kappa(c)$.
Against the time-ordered items of view $O_k$,
we compute a one-to-one partial alignment
$\pi_k^{(c)} : E_c \to O_k \cup {\emptyset}$
that maximizes total temporal overlap over matched references:
\begin{equation}
\pi_k^{(c)} = \operatorname*{arg\,max}_{\pi} \sum_{\substack{e \in E_c \\ \pi(e) \neq \emptyset}} \text{IoU}(e, \pi(e)) \quad \text{s.t.} \quad
\text{IoU}(e, \pi(e)) \geq \tau_k \;\; \forall\, e : \pi(e) \neq \emptyset
\label{eq:injective}
\end{equation}
where $\tau_k$ is minimum temporal overlap threshold. 
We then construct the fused target timeline $\hat{S}$ by taking each event's temporal span from its temporal owner and each attribute from its designated authority:
\begin{equation}
\hat{S} = \bigcup_{c} \left\{ e \ \middle|\ \forall \alpha \in \mathcal{A}, \ \tau(e) = \tau(e^{\kappa(c)}), \ 
\alpha(e) = g_{\alpha}\left( \pi_{\text{auth}(\alpha)}^{(c)}(e) \right) \right\}
\label{eq:fused_scene}
\end{equation}
where $e^{\kappa(c)}$ represents the class $c$ events of the temporal owner, and $g_{\alpha}$ is a deterministic attribute selection function 
(for instance, longest string for open-text descriptions)
that resolves cases where an annotator contributes multiple candidate values. 

\paragraph{Instruction Input Output completion format} 
Decoding a nested, multi-granularity timeline in a single pass imposes an all-or-nothing inference paradigm.
In practice, however, downstream users may already possess high-quality partial signals such as an expert human transcript or output from a specialized diarization system.
A standard end-to-end captioner cannot naturally incorporate these external priors to condition its remaining predictions, 
limiting its use as a modular component in larger audio pipelines.

\begin{figure*}[h]
\centering
\includegraphics[width=\textwidth]{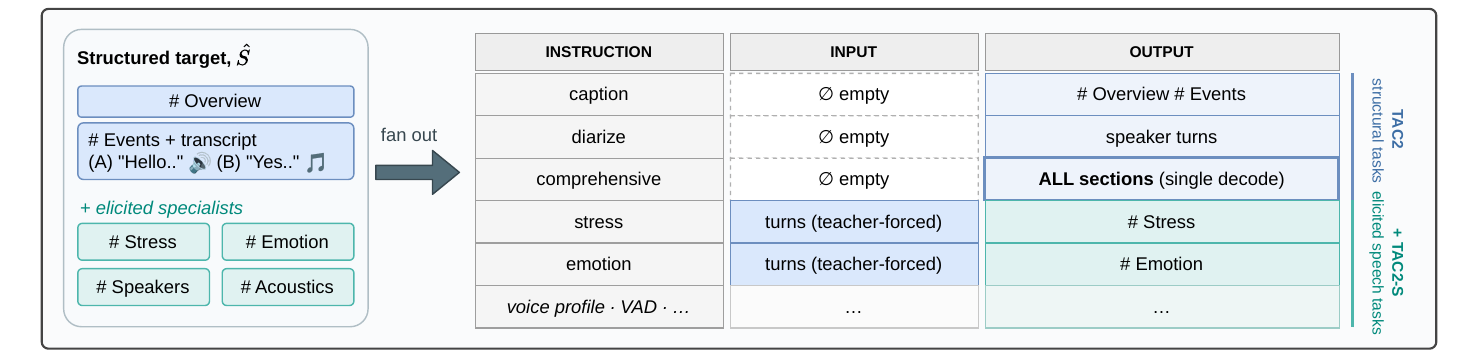}
\caption{\textbf{The IIO completion format.} 
The fused target $\hat{S}$ is decomposed into 
(\texttt{INSTRUCTION}, \texttt{INPUT}, \texttt{OUTPUT}) 
completion tasks. 
The \texttt{INSTRUCTION} specifies the task, 
\texttt{INPUT} provides optional teacher-forced context, and 
\texttt{OUTPUT} specifies the target attributes to generate. 
Structural tasks train \modelname{} (\texttt{caption}, \texttt{diarize}, and a single-decode \texttt{comprehensive} task). 
Appending elicited speech tasks (\texttt{stress}, \texttt{emotion}, \texttt{voice profile}, \texttt{acoustics}) 
seamlessly extends this format to \studentname{} without changing the training procedure. 
At inference, 
the \texttt{comprehensive} task supports single-pass generation, 
while sequencing attribute-specific tasks with preceding outputs yields a modular multi-pass cascade.}
\label{fig:iio}
\end{figure*}

We address this limitation with an \emph{Instruction-Input-Output} (IIO) completion format,
which fans out the target schema into completion tasks over the same audio
(Figure~\ref{fig:iio}). 
The \texttt{INSTRUCTION} specifies the task, 
the \texttt{INPUT} is either empty or contains teacher-forced context such as a transcript or speaker turns, 
and the \texttt{OUTPUT} is the remaining schema that we expect the model to generate. 
We train \modelname{} as a LoRA adapter $\Delta W_{\text{base}}$ on a frozen pre-trained backbone $W_0$,
with adapted parameters $\theta_{\text{base}} = W_0 + \Delta W_{\text{base}}$, 
by minimizing the completion loss over the fanned-out tasks:
\begin{equation}
\mathcal{L}_{\text{base}} = - \sum_{t=1}^{|Y_{\text{base}}|}
\log P\left( y_{t}^{\text{base}} \ \middle|\
Z_a, \ y_{<t}^{\text{base}}, \ X_{\text{inst}}, \ X_{\text{input}} ; \
W_0 + \Delta W_{\text{base}} \right)
\end{equation}
where $Z_a$ is the audio, 
$X_{\text{inst}}$ the instruction, 
$X_{\text{input}}$ the optional teacher-forced context, and 
$Y_{\text{base}}$ the target completion.

Two properties make IIO central to our approach. 
Firstly, it is \emph{extensible}: 
new attributes enter the schema as additional completion tasks without changing the training procedure.
Stage~2 (Section~\ref{sec:stage2_method}) 
exploits this property by appending elicited speech attributes to the same task fan-out
(Figure~\ref{fig:iio}, + \studentname{} rows). 
\swap{Second, each IIO task acts as a \emph{composable function} $f_\tau(Z_a, X_{\text{input}})$, that is, the model prompted with the instruction of task $\tau$.
With an empty \texttt{INPUT}, the \texttt{comprehensive} task generates the full acoustic timeline in a single call, $f_{\text{comprehensive}}(Z_a, \emptyset)$.
Alternatively, attribute-specific tasks take preceding predictions 
or externally supplied priors as \texttt{INPUT},
so calls compose into a multi-pass cascade, for example $f_{\text{stress}}\big(Z_a, f_{\text{caption}}(Z_a, \emptyset)\big)$.}
\swap{Likewise}, 
injecting a high-quality transcript from an external ASR system allows subsequent attributes to be predicted while holding the transcript fixed.
This same mechanism enables \swap{data bootstrapping} as stronger external priors become available.
We compare these inference modes in Section~\ref{sec:exp_singlepass} and detail the per-task sampling mixture in Appendix~\ref{app:singlepass}.

\subsection{Stage 2: Speech-centric Elicitation (\studentname{})}
\label{sec:stage2_method}

We next extend \modelname{}'s physical timeline with expressive speech attributes, including word stress, intonation, emotion, and speaker characteristics.
Rather than learning these capabilities from scratch at scale, we elicit paralinguistic information already encoded within the pre-trained backbone using lightweight specialists,
then distill their annotations into the unified model, \studentname{}.

\paragraph{Parameter-efficient Specialist Elicitation} 
Clean paralinguistic labels are scarce at the scale required for our in-the-wild corpus.
We therefore formulate this expansion as \emph{latent capability elicitation}.
Pre-trained models can encode capabilities internally that are not reliably expressed in their outputs, but can be surfaced through limited parameter updates~\citep{donoway2025quantifying}.
We hypothesize that the pre-trained LALM backbone similarly encodes expressive acoustic traits that are poorly exposed through its textual generation.

We test this hypothesis through layer-wise linear probing of the intermediate hidden states
$h_\ell$ of the Qwen3Omni-captioner model. 
For a binary expressive attribute $c_i \in \{0, 1\}$, 
we fit an $L_2$ regularized logistic probe $g_\ell$ at every layer $\ell \in [0, L]$ by minimizing the cross-entropy:
\begin{equation}
\min_{w_\ell, b_\ell} \; -\sum_{i=1}^{|\mathcal{D}|} 
\Big[ c_i \log\sigma(z_{\ell, i}) + (1 - c_i) \log\big(1 - \sigma(z_{\ell, i})\big) \Big] + \lambda \| w_\ell \|_2^2
\end{equation}
where 
$z_{\ell, i} = w_\ell^T h_{\ell, i} + b_\ell$ and $\sigma(z)$ is the sigmoid function.
Our analysis reveals a gap between \emph{latent decodability} and \emph{generative accessibility}.
Paralinguistic traits are strongly linearly decodable from the frozen backbone despite substantially weaker textual attribute generation \swap{(Table~\ref{tab:elicit_output})}.
Specialist adaptation leaves early- and mid-layer decodability largely unchanged, with gains concentrated in later layers.
Appendix~\ref{app:elicitation} provides the layer-wise analysis.

Motivated by this latent perception,
we perform sample-efficient acoustic elicitation. 
More concretely, instead of using large datasets to teach the model how to perceive prosody from scratch, 
we train attention-only Low-Rank Adaptation (LoRA) specialists on small gold-standard datasets ($<10$K samples).
Given the strong decodability of these attributes in the frozen backbone, 
these specialists primarily learn to expose the latent representations as explicit, structured output tags rather than learning the underlying acoustic features from scratch.

We train one LoRA specialist for each speech attribute $e\in\mathcal{E}$ (such as word stress, intonation, emotion, and voice profile) 
by adding a parameter efficient adapter $\Delta W_{\text{spec}, e}$ to the frozen base $W_0$:
\begin{equation}
\mathcal{L}_{\text{spec}, e} = - \sum_{t=1}^{|Y_e|} 
\log P\left( y_{t, e} \ \middle|\ 
Z_a, \ y_{<t, e}, \ 
X_{\text{inst}, e}, \ X_{\text{input}, e} ; \ 
W_0 + \Delta W_{\text{spec}, e} \right)
\end{equation}
where $X_{\text{input}, e}$ supplies a task-specific (teacher-forced) context, 
such as transcript or speaker list.
The context allows each specialist to focus on predicting its target attribute $Y_e$ rather than re-deriving the surrounding description.
We detail the conditioning for specialists in Appendix~\ref{sec:exp_impl}. 

Once trained, 
we hot-load the specialists as co-existing LoRAs over the frozen Qwen3Omni-captioner to form our teacher cascade, \teachername{}.
We deploy this over our in-the-wild corpus to augment the Stage~1 timelines with speech-centric, multi-attribute distillation targets.

\paragraph{Unified Student Distillation} 
We distill \teachername{} into a single unified student, \studentname{}.

Starting from the Stage~1 \modelname{} weights $\theta_{\text{base}}$, we learn a student LoRA $\Delta W_{\text{stu}}$:
\begin{equation}
\mathcal{L}_{\text{dist}} = - \sum_{t=1}^{|Y_{\text{casc}}|} 
\log P_{\text{stu}} \left( y_{t}^{\text{casc}} \ \middle|\ 
Z_a, \ y_{<t}^{\text{casc}}, \ 
X_{\text{inst}}, \ X_{\text{input}} ; \ 
\theta_{\text{base}} + \Delta W_{\text{stu}} \right)
\end{equation}
where $Y_{\text{casc}}$ denotes the enriched targets produced by \teachername{}.
The paralinguistic attributes are incorporated as additional fanned-out IIO tasks mixed with the Stage~1 objectives,
allowing \studentname{} to learn multi-granularity completion across the distillation target.

%% file: sec/4_experiment.tex
\section{Experiments}
\label{sec:experiments}
A good describer should encode attributes relevant for both reasoning and generation in its text output. Here, we evaluate our trained captioners in these two settings: 
\textit{audio understanding}, 
by cascading \studentname{} with a frozen text-only LLM which answers questions based on \studentname{}'s generated description, and 
\textit{audio generation}, 
by using its structured descriptions as paired supervision for text-to-audio generation.

\subsection{Audio Understanding: validating schema fidelity}
\label{sec:exp_understanding}

\begin{figure}[h]
\centering
\includegraphics[width=0.97\columnwidth]{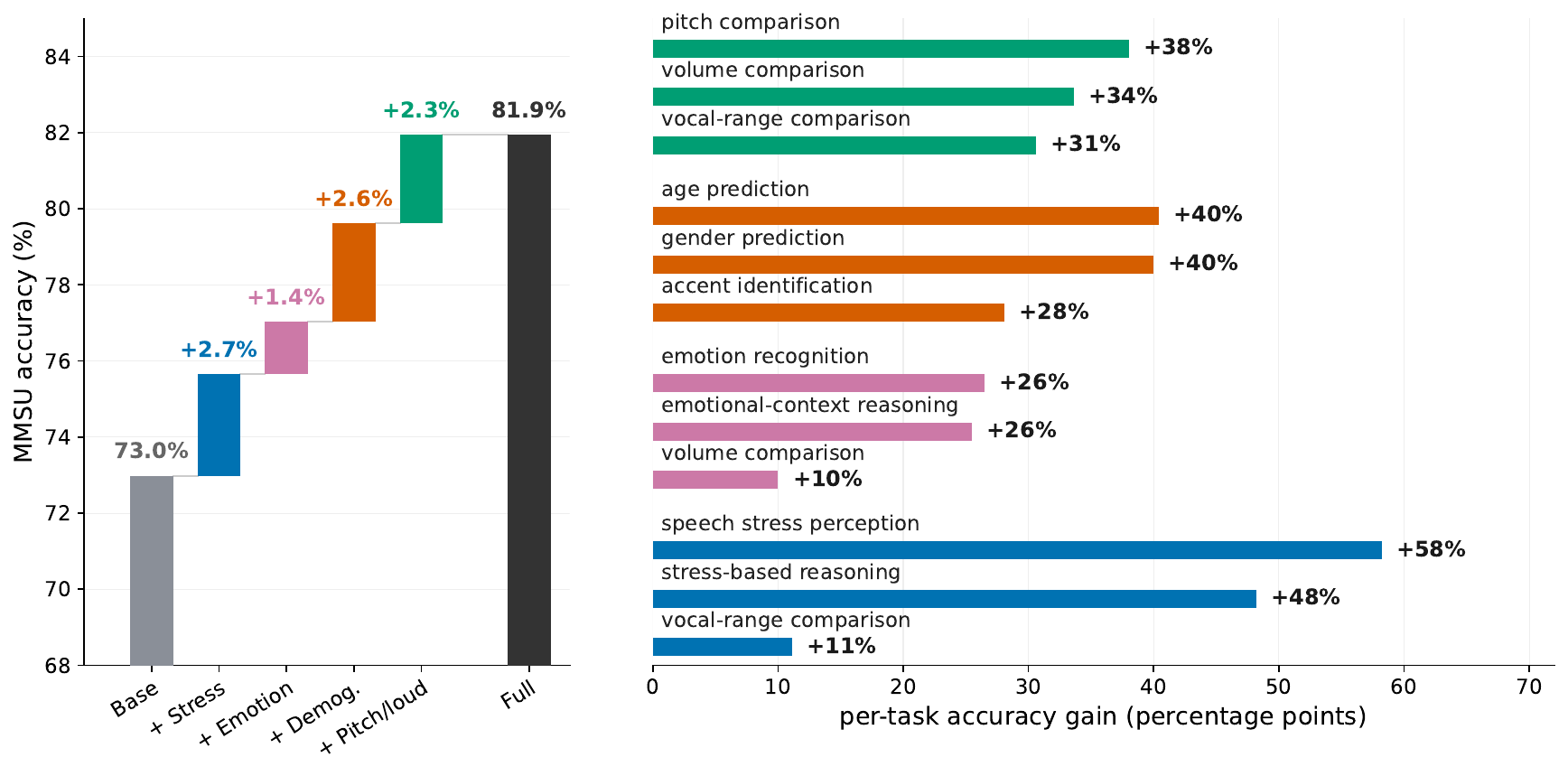}
\caption{
\textbf{Impact of individual IIO tasks.} 
\textbf{(left)} Building \studentname{}'s description one attribute at a time, 
from a transcript-only base (who spoke, the words, and their timestamps). 
\textbf{(right)} For each attribute, the top-3 it lifts most, the gains land on the tasks that attribute governs and leave the transcript carried semantics tasks flat. 
Bar colour marks the added attribute 
(blue: stress and intonation, 
pink: emotion, 
orange: voice profile(demographics), 
green: pitch and loudness).}
\label{fig:buildup}
\end{figure}

Under the \emph{Describe-then-Reason (DtR)} protocol~\citep{tac}, 
\studentname{} converts audio into a structured acoustic schema that a frozen text-only LLM uses to answer multiple-choice questions.
This helps us assess whether the schema preserves the perceptual information required for downstream reasoning.
We evaluate on 
MMSU~\citep{mmsu}, 
the general audio benchmarks MMAR~\citep{mmar} and MMAU~\citep{mmau}, and 
MSU-Bench~\citep{msubench}.
\studentname{} achieves an 81.9 overall score on MMSU, outperforming existing LALMs and nearing \teachername{}, the teacher cascade (82.9).
The gain is concentrated in perception (Table~\ref{tab:final_unified_results}): 
\studentname{} reaches 77.8 on MMSU perception, 
compared with 68.7 for Gemini-3.7 Flash, 63.8 for TAC, and 63.5 for Qwen3Omni-Captioner,
while retaining strong reasoning performance (86.4).
The attribute buildup in Figure~\ref{fig:buildup} further localizes these gains, 
whereby adding individual schema attributes preferentially improves their  associated perception tasks while leaving transcript-driven semantic tasks comparatively stable.
This advantage is specific to fine-grained acoustic representation rather than general audio QnA alone.
Frontier OmniLLMs remain weaker on fine-grained acoustics, timing, loudness, and phonetics, whether operating directly on audio or under the same \emph{DtR} protocol (Table~\ref{tab:mmsu-ablations}, right).
Beyond MMSU, \studentname{} remains competitive on the broader MMAR and MMAU benchmarks, reaching 68.0 and 75.6, respectively.
On the speaker-centric MSU-Bench, 
it achieves 82.2 overall 
(79.2 Tier~1; 86.6 Tier~2),
outperforming the baselines and \teachername{}
(Table~\ref{tab:msubench}).
\footnote{For $\ge$ 30s clips, we chunk and use MOSS-0.9B \citep{ai2026moss} solely for cross-chunk speaker consistency.}

\begin{table*}[t]
\centering
\setlength{\tabcolsep}{2.5pt}
\renewcommand{\arraystretch}{1.15}
\caption{
\textbf{Spoken and general audio understanding.} 
Models in \textbf{DtR \cmark} rows emit a text schema as the description, which is read by a frozen text-only reasoner.
Each score reflects the fidelity and coverage of audio perception in its generated description. 
}
\resizebox{\textwidth}{!}{%
\begin{tabular}{l c ccc cccc cccc c}
\toprule

\multirow{2}{*}{\textbf{Model}} 
& \multirow{2}{*}{\textbf{DtR}} 
& \multicolumn{3}{c}{\textbf{MMSU}} 
& \multicolumn{4}{c}{\textbf{MMAR}} 
& \multicolumn{4}{c}{\textbf{MMAU}} 
& \multirow{2}{*}{\textbf{Avg.}} \\

\cmidrule(lr){3-5} \cmidrule(lr){6-9} \cmidrule(lr){10-13}

& & Perception & Reasoning & Overall 
& Speech & Sound & Music & Overall
& Speech & Sound & Music & Overall
& \\

\midrule
Kimi-Audio
& \xmark
& 44.8 & 75.7 & 59.8
& 58.5 & 49.7 & 33.0 & 48.0
& 62.2 & 75.7 & 66.8 & 68.2
& 58.7 \\

Qwen2.5-Omni
& \xmark
& 42.7 & 77.6 & 58.1
& 59.9 & 58.8 & 40.8 & 56.7
& 70.6 & \underline{78.1} & 65.9 & 71.5
& 62.1 \\

StepAudio2
& \xmark
& 42.9 & 73.2 & 57.6
& 61.2 & 54.6 & 42.2 & 56.8
& 68.2 & \textbf{79.3} & 68.4 & 72.7
& 61.9 \\

UAS-Audio
& \xmark
& 55.7 & 77.4 & 66.2
& 66.0 & 58.8 & 45.2 & 60.1
& 67.0 & 70.0 & 71.3 & 69.4
& 65.2 \\

Gemini-3.7 Flash 
& \cmark
& 68.7 & 84.4 & 76.3
& 82.7 & 51.5 & \underline{56.8} & 69.5
& 80.5 & 66.1 & 67.1 & 71.2
& 72.3 \\

Qwen3Omni-Capt. 
& \cmark
& 63.5 & 83.4 & 73.1
& 77.6 & \underline{59.4} & 54.4 & 69.2
& \textbf{84.1} & 73.0 & 68.0 & \underline{75.0}
& 72.4 \\

TAC 
& \cmark
& 63.8 & 84.8 & 74.0
& \underline{83.7} & \textbf{62.4} & \textbf{57.3} & \textbf{71.1}
& 82.0 & 69.4 & \underline{69.8} & 73.7
& 72.9 \\

\midrule
\teachername{} 
& \cmark
& \textbf{79.0} & \textbf{87.1} & \textbf{82.9}
& \textbf{84.0} & 56.4 & 52.9 & \underline{70.3}
& 80.5 & 71.8 & 68.3 & 73.5
& \textbf{75.6} \\

\studentname{} 
& \cmark
& \underline{77.8} & \underline{86.4} & \underline{81.9}
& 79.6 & 53.3 & 51.0 & 68.0
& \underline{82.3} & 73.9 & \underline{70.7} & \textbf{75.6}
& \underline{75.2} \\

\bottomrule
\end{tabular}%
}
\label{tab:final_unified_results}
\end{table*}

\begin{table}[t]
\centering
\caption{
\textbf{Inference and Reasoner Robustness.}
\textbf{Left:}
\studentname{} schemas under single-pass (SP) vs.\ multi-pass (MP) comprehensive decodes across text-only reasoners (top), and the MMSU tasks where MP helps most (bottom).
\textbf{Right:} Perception fidelity.
\studentname{} uses describe-then-reason; OmniLLM baselines process raw audio directly (left radar) or act as describers (right radar).
\textcolor{itadblue}{\rule{1.2ex}{1.2ex}}~\studentname{} (ours),
\textcolor{geminiorange}{\rule{1.2ex}{1.2ex}}~Gemini 3.7,
\textcolor{qwenaqua}{\rule{1.2ex}{1.2ex}}~Qwen3Omni
}
\label{tab:mmsu-ablations}
\begin{minipage}[c]{0.26\textwidth}
\centering
\footnotesize
\setlength{\tabcolsep}{3pt}
\begin{tabular}{clcc}
\toprule
\multicolumn{2}{l}{Text-only reasoner} & SP & MP \\
\midrule
\multirow{3}{*}{\rotatebox[origin=c]{90}{\studentname{}}}
& $\rightarrow$ Gemini3 & 81.3 & 81.9 \\
& $\rightarrow$ Qwen3-Next & 78.8 & 79.7 \\
& $\rightarrow$ Qwen3.8  & 80.5 & 80.9 \\
\bottomrule
\end{tabular}

\vspace{1em}

{
\begin{tabular}{lcc}
\toprule
Tasks (MMSU)  & SP & MP \\
\midrule
Pitch comparison   & 63.0 & 75.9 \\
Vocal-range comp.\ & 60.2 & 72.2 \\
Emotional context  & 43.6 & 55.5 \\
Accent ident.          & 59.6 & 69.3 \\
\bottomrule
\end{tabular}
}
\end{minipage}%
\hfill
\begin{minipage}[c]{0.73\textwidth}
\centering
\includegraphics[width=0.9\linewidth]{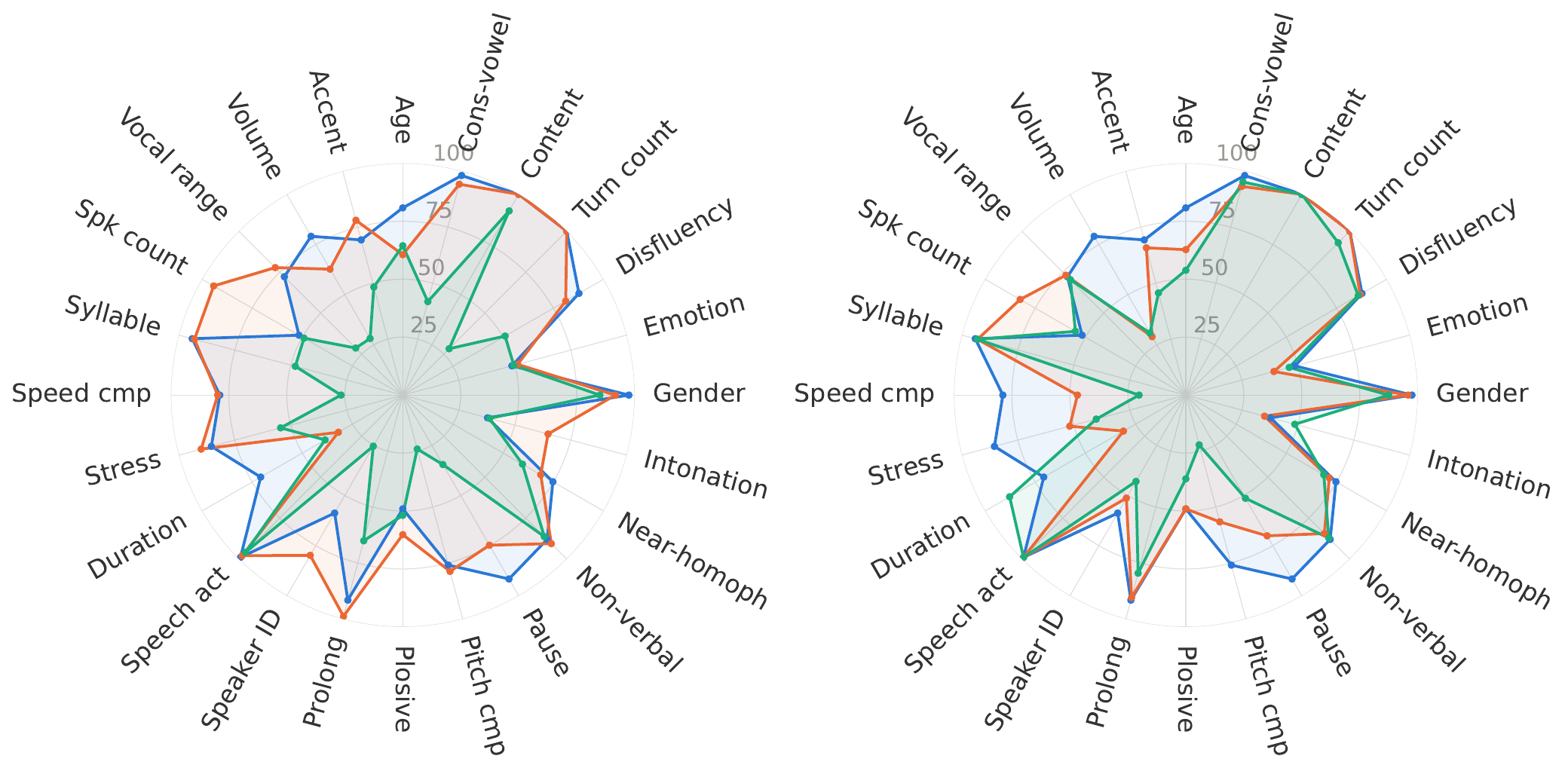}
\end{minipage}
\end{table}

\begin{table}[t]
\centering
\caption{
\textbf{Speaker-centric understanding on MSU-Bench.} 
Evaluated on conversational multi-speaker scenarios in English/Mandarin.
\textbf{Left:} Consistent with Table~\ref{tab:final_unified_results}, \studentname{} uses a text reasoner.
\textbf{Right:} Tier 1 (Identification) task accuracy.
(AI: accent identification, 
AR: age recognition, 
GI: gender identification, 
ER: emotion recognition, 
SP: speaker profiling,                                                       
SR: speaker retrieval, 
SVS: speaker-specific viewpoint summarization, 
SC: speaker counting, 
SV: speaker verification)
OmniLLM baselines process raw audio directly (left radar) or act as describers (right radar). \studentname{}'s trace (blue) is superior across both, demonstrating robust speaker-centric perception.
\textcolor{itadblue}{\rule{1.2ex}{1.2ex}}~\studentname{} (ours),
\textcolor{geminiorange}{\rule{1.2ex}{1.2ex}}~Gemini 3.7,
\textcolor{qwenaqua}{\rule{1.2ex}{1.2ex}}~Qwen3Omni
}
\label{tab:msubench}
\begin{minipage}[c]{0.32\textwidth}
\centering
\footnotesize
\setlength{\tabcolsep}{3pt}
\begin{tabular}{lccc}
\toprule
\textbf{System} & \textbf{T1} & \textbf{T2} & \textbf{Avg} \\
\midrule
Kimi-Audio      & 41.0 & 47.0 & 43.0 \\
Qwen2.5-Omni    & 19.0 & 21.0 & 19.0 \\
StepAudio2      & 44.0 & 46.0 & 44.0 \\
AudioFlamingo-3 & 40.0 & 38.0 & 39.0 \\
MiMoAudio       & 52.0 & 64.0 & 56.0 \\
Gemini-3.7 Flash & 70.7 & 84.8 & 76.4 \\
Qwen3-Omni       & 69.9 & 83.2 & 75.3 \\
\midrule
\teachername{}     & \underline{78.2} & \textbf{87.3} & \underline{81.9} \\
\studentname{}   & \textbf{79.2} & \underline{86.6} & \textbf{82.2} \\
\bottomrule
\end{tabular}
\end{minipage}%
\hfill
\begin{minipage}[c]{0.64\textwidth}
\centering
\includegraphics[width=\linewidth]{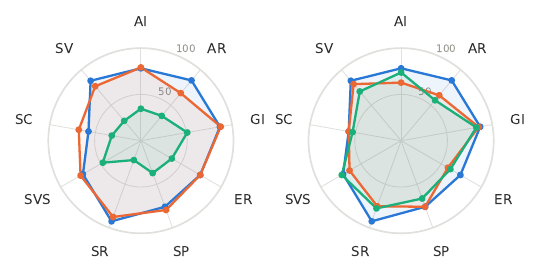}
\end{minipage}
\end{table}

\paragraph{Inference and Reasoner Robustness.}
\label{sec:exp_singlepass}
We ablate IIO decoding and the text reasoner in Table~\ref{tab:mmsu-ablations}, left.
Collapsing multi-pass (MP) inference, where each task conditions on the accumulated description, into a single (SP) decode incurs a modest MMSU drop ($81.9\to81.3$).
However, MP yields larger gains on perceptual tasks such as 
pitch ($+12.9$),
vocal range ($+12.0$), and
accent identification ($+9.7$).
Holding the \studentname{} schema fixed and swapping the reasoner from Gemini3.5-Flash with Qwen3-Next or Qwen3.8-Omni 
retains similar MP performance (79.7 and 80.9, respectively). 
In contrast to TAC~\citep{tac}, where scaling the reasoner improved results, this relative robustness to reasoner choice suggests that \studentname{}'s detailed schema already exposes much of the perceptual information needed for downstream reasoning.
Together, these results show that 
\swap{composing IIO functions} can improve fine-grained acoustic inference through intermediate conditioning, 
while the resulting schema remains effective across different text-only reasoners.

\subsection{Generation: The Description as Paired Supervision for Audio Synthesis}
\label{sec:exp_generative}
\begin{figure*}[h]
\centering
\includegraphics[width=0.97\textwidth]{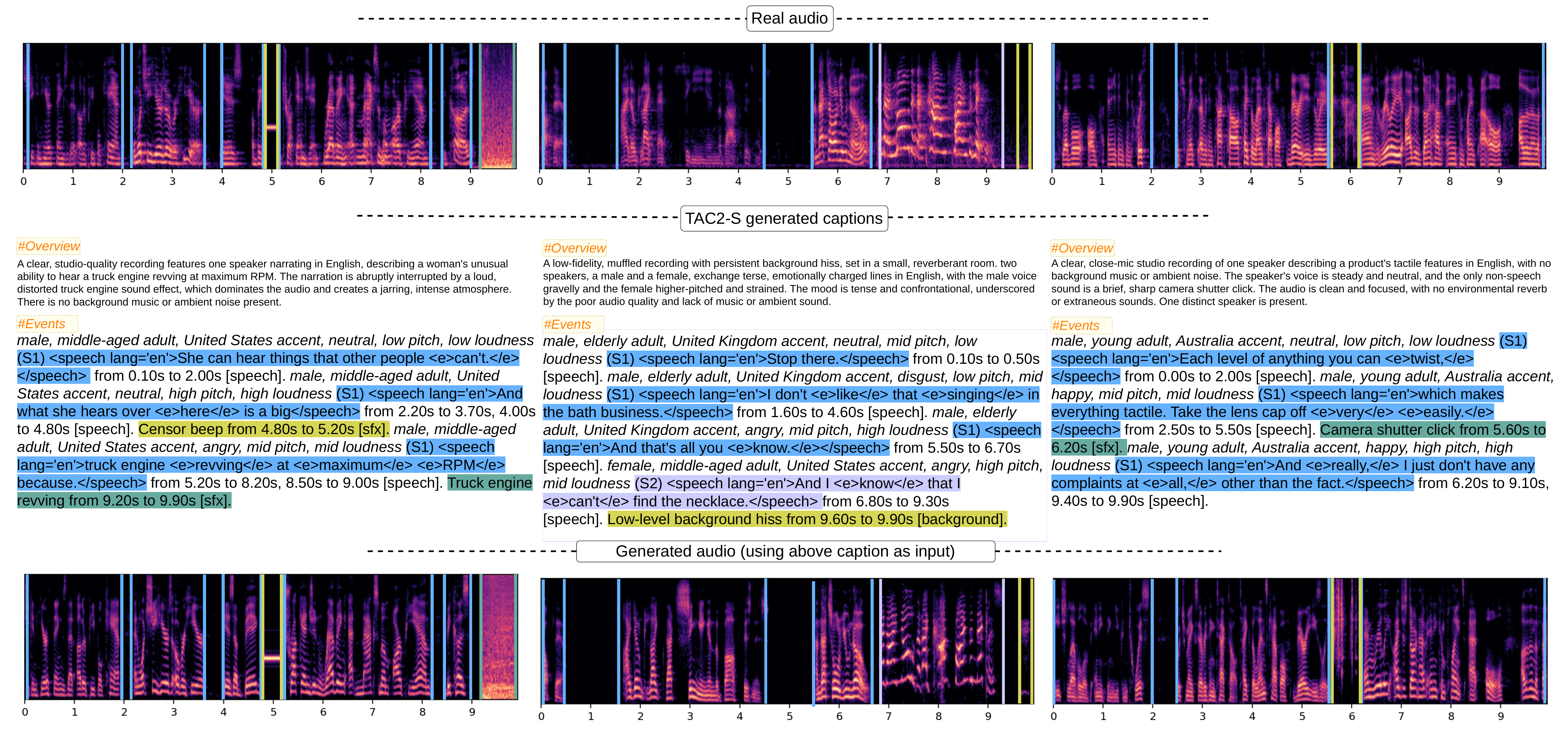}
\caption{
\textbf{\studentname{}-conditioned reconstruction.} 
We caption real audio (top) with \studentname{} (middle), 
then condition the trained text-to-audio generator on it to synthesize new audio (bottom). 
The reconstruction reproduces speech and non-speech events, 
including background elements like a \texttt{censor beep}, \texttt{engine rev}, \texttt{camera click}, at times aligned with the real audio.
}
\label{fig:t2a2t}
\end{figure*}

Our second validation tests whether the structured perceptual representation is sufficiently grounded to be \emph{renderable} back into audio.
We use \studentname{} to relabel in-the-wild audio at scale, 
producing structured timestamped captions as paired supervision for a text-to-audio diffusion model.
If the generated schemas faithfully capture the underlying scene, 
conditioning on its event timeline should reconstruct both 
\emph{what} occurs and \emph{when} it occurs (Figure~\ref{fig:t2a2t}).

  \begin{table}[t]\centering
  \caption{
  \textbf{
  Structured captions generated by \studentname{} provide renderable supervision for audio synthesis.}
  Temporal-grounding cells report \emph{with}$\mid$\emph{without} input timestamps.
  PE and FLAM measure mean localizer-peak offset in seconds; PE-AV is a text-free frame-wise semantic distance between original and reconstructed audio (real-vs.-real $=0$).
  Gemini-3-Pro judgments are on a $1$--$5$ scale.
  }
  \label{tab:tta_reconstruction}
  \resizebox{\columnwidth}{!}{%
  \begin{tabular}{l ccccccc ccc cc}
  \toprule
   & \multicolumn{7}{c}{\textbf{\studentname{}-as-judge}} & \multicolumn{3}{c}{\textbf{Temporal grounding}} & \multicolumn{2}{c}{\textbf{Gemini-3-Pro judge}} \\
  \cmidrule(lr){2-8}\cmidrule(lr){9-11}\cmidrule(lr){12-13}
  System & Gen. & Age & Acc. & Emo. & Ev.R. & Sp.R. & WER & PE & FLAM & PE-AV & vs.\ real & vs.\ cap. \\
   & $\uparrow$ & $\uparrow$ & $\uparrow$ & $\uparrow$ & $\uparrow$ & $\uparrow$ & $\downarrow$ &
   $\downarrow$ & $\downarrow$ & $\downarrow$
   &
   $\uparrow$ & $\uparrow$
   \\
  \midrule
  PicoAudio2~\citep{picoaudio2} & 0.44 & 0.35 & 0.24 & 0.28 & 0.27 & 0.09 & 1.15 & 2.47$\mid$2.77 & 3.16$\mid$3.50 & 0.52$\mid$0.51 & 1.39 & 1.61 \\
  MiDashengLM-Gen~\citep{midashenglmgen} & 0.54 & 0.36 & 0.21 & 0.36 & 0.33 & 0.09 & 1.39 & 3.57$\mid$--- & 3.78$\mid$--- & 0.44$\mid$--- & 1.54 & 1.60 \\
  Seed Audio 1.0~\citep{seedaudio} & \underline{0.94} & \textbf{0.89} & \textbf{0.71} & \underline{0.74} & \underline{0.78} & 0.90 & \underline{0.14} & 2.16$\mid$--- & 2.62$\mid$--- & \underline{0.35}$\mid$--- & \underline{3.53} & \underline{4.05} \\
  \midrule
  Ours (\modelname{})    & --- & --- & --- & --- & \textbf{0.80} & \textbf{0.94} & \textbf{0.12} & \underline{1.67}$\mid$2.47 & \textbf{2.13}$\mid$2.77 & \underline{0.35}$\mid$0.35 & 3.38 & 4.01 \\
  Ours (\studentname{})  & \textbf{0.99} & \underline{0.84} & \underline{0.69} & \textbf{0.77} & \textbf{0.80} & \underline{0.93} & \underline{0.14} & \textbf{1.54}$\mid$2.93 & \underline{2.14}$\mid$3.16 & \textbf{0.30}$\mid$0.33 & \textbf{3.95} & \textbf{4.52} \\
  \bottomrule
  \end{tabular}}
  \end{table}
  
We condition the generator on the schema of held-out clips and localize events in the synthesized audio using two frame-level audio-text localizers,
FLAM~\citep{flam} and PE-A-Frame~\citep{peaframe}. 
Temporal error is measured as the mean absolute offset between localizer peaks in the original and reconstructed audio
(Table~\ref{tab:tta_reconstruction}). 
Our generator provides the strongest temporal grounding,
achieving $1.54$s PE and $2.14$s FLAM error, 
compared with $2.47$s and $3.16$s for the explicitly timestamp-conditioned PicoAudio2~\citep{picoaudio2}.
Removing timestamps from the \studentname{} captions substantially degrades localization 
($1.54\to2.93$s PE; $2.14\to3.16$s FLAM), 
thereby demonstrating that the generator uses the predicted temporal structure.

Furthermore, we compare the recovered fields against the conditioning description using \studentname{}-as-judge.
The reconstructed audio closely preserves speech content and speaker attributes, 
reaching $0.99$ gender accuracy, 
$0.84$ age accuracy, 
$0.69$ accent accuracy, 
$0.77$ emotion accuracy, 
$0.93$ speech recall, and $0.14$ WER.
An audio-native Gemini-3-Pro judge rates our audio highest for both resemblance to the original ($3.95$) and adherence to the caption ($4.52$). 
We match the strongest commercial baseline, Seed Audio~1.0~\citep{seedaudio}, on speaker attributes,
\swap{while Seed Audio~1.0 trails our generator on event timing and on both judgments.}
Additional details are in Appendix~\ref{tta_model}.

\begin{figure*}[h]
\centering
\includegraphics[width=0.95\textwidth]{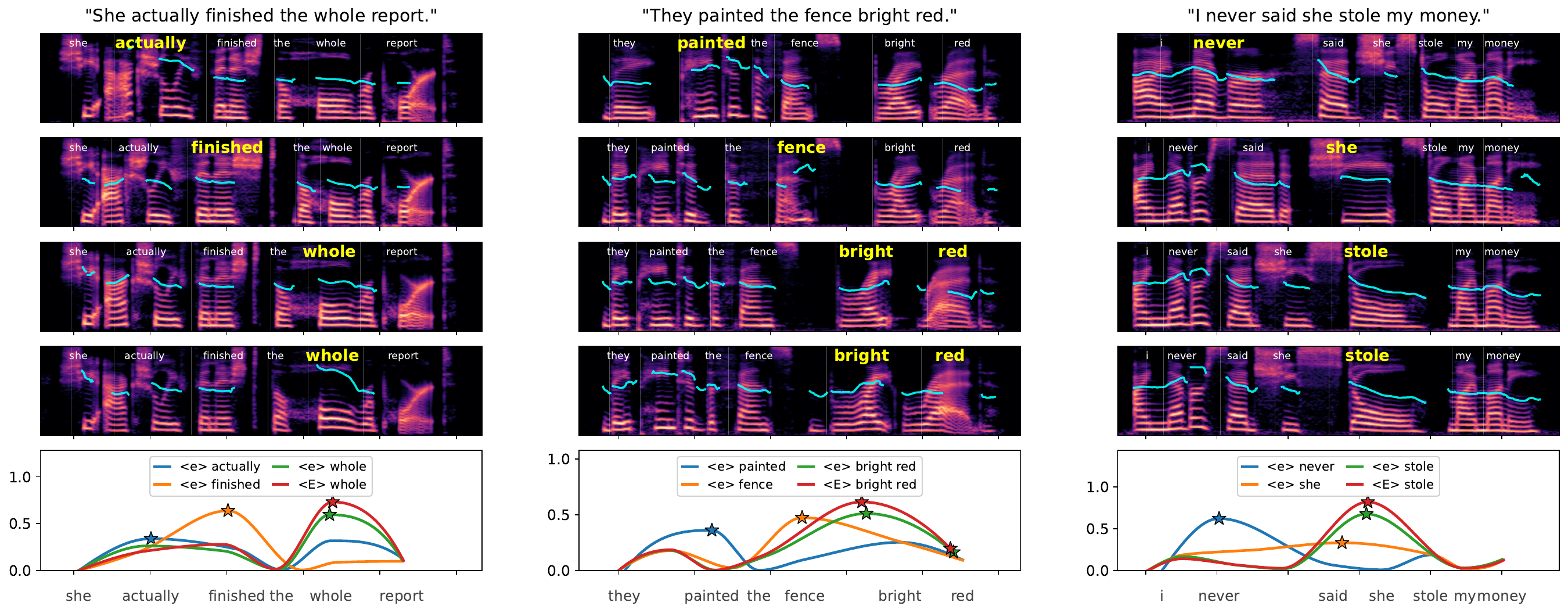}
\caption{
\textbf{Word-level emphasis as a renderable control channel.} 
Each column synthesizes the same sentence while varying the emphasized word in the \studentname{} caption (yellow).
The bottom two rows increase emphasis from mild (\texttt{<e>}) to strong (\texttt{<E>}) while the cyan curve shows the F0 contour.
Bottom panels show WhiStress~\citep{whistress} stress probability over time, with $\star$ marking the commanded word.
The detected stress peak follows the targeted word and scales with emphasis intensity, demonstrating zero-shot emphasis control without a dedicated stress-control model.
}
\label{fig:emphasis_control}
\end{figure*}

In Figure~\ref{fig:emphasis_control}, we demonstrate that \studentname{}'s structured schema exposes explicit attributes as generation-time controls. 
For a fixed sentence, we vary only the emphasized word and its intensity in the \studentname{}' caption. 
Without training a dedicated stress-control model, 
the synthesized stress peak follows the commanded word and increases with the requested emphasis level. Thus, paralinguistic attributes elicited for perception also become explicit control channels when the same representation is used as supervision for generation.

%% file: sec/5_conclusion.tex
\section{Conclusion}

We introduce \swap{\modelname{}} and its speech-centric extension, \studentname{}, 
for constructing queryable, temporally grounded acoustic representations from in-the-wild audio.
Our framework bootstraps granular, speaker-attributed timelines at scale through multi-view weak supervision, 
models them \swap{as composable functions through the} Instruction-Input-Output (IIO) completion format, 
and extends them with paralinguistic attributes through sample-efficient latent capability elicitation.
The resulting unified model provides a structured interface between perception, reasoning, and generation.
Paired with frozen text-only LLMs, it enables strong spoken audio understanding.
Used as supervision for text-to-audio generation, it supports faithful reconstruction and explicit control over event timing, speaker attributes, and word-level emphasis.
Overall, our results demonstrate that a high-fidelity, structured, temporally grounded acoustic timeline can function as both 
an \emph{interpretable} reasoning-ready representation of audio and 
a \emph{renderable} control interface for generation.

%% file: sec/6_reprod.tex
\newpage

\section*{AI use statement}
The authors used generative AI to assist with language editing of the manuscript. 
We have reviewed and take responsibility for the final content of this work, including text produced with the aid of generative AI.
All scientific content and conclusions were produced by the authors.

\section*{Reproducibility statement}
Our two-stage method is described in Section~\ref{sec:methodology}, 
with architecture and training details (backbone, LoRA rank and adapted modules, optimizer, step counts for \modelname{} and \studentname{}) in Appendix~\ref{sec:exp_impl}.
The time-registered IoU matching is formalized in Eq.~\ref{eq:iou}--\ref{eq:fused_scene} (Figure~\ref{fig:iou_matching}).
The IIO format, and the per-task mixture and inference regimes in Section~\ref{sec:stage1_method}, Figure~\ref{fig:iio}, and Appendix~\ref{app:singlepass};
The specialist elicitation, probing setup, and per-attribute gold datasets are detailed in Section~\ref{sec:stage2_method} and Appendix~\ref{app:elicitation}.
The text-to-audio validation (generator, evaluation set, metrics, baselines, and judge) and all evaluation protocols and judge prompts are provided in Appendices~\ref{tta_model} and~\ref{sec:prompts}.
Our backbone (Qwen3-Omni-Captioner), 
the specialist gold corpora (IEMOCAP~\citep{iemocap}, RAVDESS~\citep{ravdess}, MELD~\citep{meld}, Common Voice~\citep{commonvoice}), and 
all benchmarks are publicly available.
Our in-the-wild corpus uses proprietary licensed audio sources, instead we provide complete time-registered IoU matching to reproduce our results with alternative audio sources.

%% file: sec/appendix.tex
\section*{Summary of Appendix}
\begin{itemize}[nosep]
    \item Section~\ref{app:limitations}: Limitations and Future Work
    \item Section~\ref{sec:exp_impl}: Implementation Details
    \item Section~\ref{app:elicitation}: Elicitation: Per-Layer Probing
    \item Section~\ref{app:singlepass}: IIO Tasks
    \item Section~\ref{app:localization}: Segment- and pause-level localization
    \item \swap{Section~\ref{app:longaudio}: Long-form audio understanding}
    \item Section~\ref{app:tac2v}: Extending the Describer to Audio-Visual Reasoning
    \item Section~\ref{tta_model}: Text-to-Audio Model and Temporal Grounding Setup
    \item Section~\ref{sec:prompts}: Prompts
\end{itemize}

\section{Limitations and Future Work}
\label{app:limitations}

Writing perception as an explicit textual description makes \studentname{}'s errors readily interpretable, 
tracing back to either an incorrect recognition of an existing attribute or a lack of schema coverage.
Currently, \studentname{}'s schema prioritizes spoken attributes over non-speech and music events, 
a limitation reflected in benchmarks such as MMAR and MMAU. 
Interestingly, our current schema format enforces a temporal resolution tradeoff. 
Repeated events are merged into one span 
(for example, \texttt{<speech>...</speech> from 1.4s to 3.8s, 5.4s to 6.7s}) 
with comma-separated timestamps.
While this fine temporal granularity helps \studentname{} act as a crucial reference for tasks like pause detection, disfluency detection, duration estimation etc. (see Table \ref{tab:mmsu-ablations} (right), and Appendix~\ref{app:localization}), 
it occasionally merges turns when consecutive voices are similar, causing the model to under-count unique speakers.

It is important to note that these bottlenecks reflect coverage limits rather than our architectural constraints. 
Our IIO proposal is inherently extensible, 
and new perceptual axes as well as upgraded teachers can be seamlessly integrated via teacher-forced bootstrapping. 
More broadly, human-designed schemas will eventually hit a coverage ceiling, oversight leading to bottleneck. 
In future, we aim to let descriptions evolve autonomously via reinforcement learning, 
using frozen text reasoners to reward the extraction of task-demanded cues.

\section{Implementation details}
\label{sec:exp_impl}
\modelname{} and the individual speech specialists are independently trained as rank-512, attention-only LoRA adapters over the same frozen Qwen3-Omni-30B captioner. At inference, the specialists can therefore be hot-loaded over this shared backbone to form the modular \teachername{} cascade. The unified \studentname{} is instead trained as an additional LoRA over the Stage~1 \modelname{} weights.

\textbf{Stage 1.}
We train \modelname{} on approximately two million licensed, in-the-wild audio clips, each up to 30 seconds long.
Each clip is paired with a fused target timeline constructed using the multi-view fusion procedure described in Section~\ref{sec:stage1_method}.
We combine $K=3$ specialized weak teachers:
(1) a TAC-style sound event detector~\citep{tac} for non-speech event spans and types;
\swap{(2)} a TAC-inspired diarizer~\citep{tac} for speech turns and speaker identities; and (3) the Qwen3-Omni captioner~\citep{qwen3omni} for transcripts, voice descriptions, and scene overviews.
We additionally apply wav2vec2-based~\citep{wav2vec2} forced alignment to obtain word-level timestamps.
The per-type overlap gate $\tau_k$ of Eq.~\ref{eq:injective} is set to $0.2$~IoU for sound-effect events and 
$0$ for speech and other non-speech events (rely on the order-preserving alignment with a midpoint-proximity fallback).
A separate $0.5$~IoU gate de-duplicates overlapping event descriptions.
We selected these values by manual inspection of the merged outputs on a small validation set of clips.
The resulting targets are decomposed into the IIO tasks described in Section~\ref{sec:stage1_method},
with training samples weighted proportionally to their speaker count. 
We optimize \modelname{} \swap{with AdamW} for 10,000 steps with a learning rate of $1.6\times10^{-3}$, a 1,000-step linear warmup, and cosine decay.

\textbf{Stage 2.}
We train each speech specialist on fewer than 10,000 labeled examples, 
using task-specific teacher-forced context so that the adapter predicts only its designated paralinguistic attribute.
For word stress and intonation, we use the multilingual A560E dataset~\citep{wei2026speech}, spanning 10 languages, 
and are conditioned on the transcript, incorporating per-word relative pitch extracted via \textit{torchcrepe}~\citep{Morrison_torchcrepe_2023}. 
For emotion, we use IEMOCAP~\citep{iemocap}, RAVDESS~\citep{ravdess}, and MELD~\citep{meld}, predicting utterance-level emotion directly from the audio.
For voice profile (gender, age, and accent), we use silver labels obtained from a per-speaker Qwen3-Omni probe over the licensed corpus, supplemented with a balanced Common Voice~\citep{commonvoice} accent shard to mitigate the US-accent skew of the silver labels, 
conditioned on the caption. 
Finally, relative pitch and loudness are derived from DSP measurements (CREPE F0 and RMS energy) ranked within each clip and conditioned on the temporal event list.

Once trained, these specialists are concurrently hot-loaded \swap{alongside \modelname{} over the shared frozen backbone} to form \teachername{}, 
which is subsequently used to re-annotate the complete Stage 1 corpus. 
A two-pass LLM rewriting step, using Qwen3-Next-80B, then rectifies the overview with the per-event descriptors from the hot-loaded specialists, for example to keep speaker count, gender, and accent consistent.

We initialize \studentname{} from the Stage~1 \modelname{} weights and train an additional rank-512, attention-only LoRA on the enriched corpus. The elicited speech attributes are introduced as additional IIO tasks and mixed with the original Stage~1 objectives, preserving the same IIO fan-out used by \modelname{}.

We train \studentname{} for 2,000 steps \swap{using the same optimizer and learning rate as \modelname{}}, with a 100-step linear warmup and cosine decay. The resulting model consolidates the modular predictions of \teachername{} into a single adapter while retaining the structural capabilities learned by \modelname{}.

\section{Elicitation: Per-Layer Probing}
\label{app:elicitation}

Section~\ref{sec:stage2_method} hypothesizes that the Qwen3-Omni backbone already encodes paralinguistic information internally, while our attention-only LoRA specialists primarily improve its accessibility to late-stage generation rather than learning new perceptual features. We examine this hypothesis through layer-wise linear probing before and after specialist adaptation.

\paragraph{Per-layer probe.}
For each paralinguistic attribute, we fit an $L_2$-regularized logistic probe independently at each of the model's 49 hidden layers. Representations are extracted at the final token position corresponding to the attribute's generation step. We report 5-fold cross-validated AUC, with macro one-vs-rest averaging for multiclass attributes such as accent.

Three consistent patterns emerge (Table~\ref{tab:probe}).
First, every evaluated trait exhibits high linear decodability within the frozen backbone (peak AUCs between $0.89$ and $0.99$). Second, across the early and middle transformer blocks (depths 0--24), the representation quality of the backbone and the specialist are virtually identical (e.g., word stress AUC is $0.88$ for the base versus $0.89$ for the specialist), confirming that the adapter preserves early acoustic processing. Third, the adapter's structural impact is strictly localized to the late layers (36--48), where it maximizes performance within the available headroom. This relative gain is most pronounced for relational traits that the base model struggles to articulate (e.g., relative loudness and pitch, yielding a $+0.10$ to $+0.12$ AUC lift). For traits already near the performance ceiling internally (e.g., accent at $+0.01$ lift), the specialist functions primarily to calibrate the final output distribution rather than enhance latent separability.

\begin{table}[t]
\centering
\caption{The backbone (Qwen3Omni-captioner) possesses latent perception; the specialist isolates late-stage readouts.
\emph{Early}/\emph{Late}: mean AUC over depths 0--24 / 36--48, backbone vs.\ specialist.
\emph{Lift}: the specialist's late-stage gain.}
\label{tab:probe}
\resizebox{\columnwidth}{!}{%
\begin{tabular}{lcccc}
\toprule
Trait (granularity) & Backbone peak & Early (base / spec) & Late (base $\to$ spec) & Lift \\
\midrule
Relative loudness (segment) & 0.89 & 0.79 / 0.81 & 0.87 $\to$ 0.99 & $+0.12$ \\
Relative pitch (segment)    & 0.90 & 0.80 / 0.82 & 0.89 $\to$ 0.99 & $+0.10$ \\
Word stress (word)          & 0.93 & 0.88 / 0.89 & 0.91 $\to$ 0.99 & $+0.08$ \\
Accent (speaker)            & 0.99 & 0.97 / 0.97 & 0.99 $\to$ 1.00 & $+0.01$ \\
\bottomrule
\end{tabular}}
\end{table}

\paragraph{Mitigating output collapse and bias.} High internal decodability does not inherently translate to accurate generation. Post-training alignment heavily biases the decoding trajectory, causing the base model's textual outputs to exhibit mode collapse toward majority classes. Table~\ref{tab:elicit_output} contrasts the generated attribute accuracy of the frozen backbone against the specialist on a ground-truth evaluation set.

The baseline generation errors are highly systematic. Accent predictions overwhelmingly default to ``United States'' (emitted for $42\%$ of samples compared to a $9\%$ true prior), and emotion predictions degenerate to ``neutral'' ($76\%$ of samples, despite ``neutral'' not appearing as a valid ground-truth class). Concurrently, word stress exhibits a severe false-positive rate ($1.5$ stressed words per clip at $0.21$ precision). 

The specialist adapter effectively mitigates this generation bias, substantially recovering the true attribute distributions without altering the underlying perceptual manifold (consistent with the identical early-layer AUCs). Furthermore, the specialist's residual accent errors primarily occur between geographically adjacent regions (e.g., Canada $\to$ United States, Ireland $\to$ United Kingdom, New Zealand $\to$ Australia, Kenya $\to$ Nigeria). Consequently, strict 13-class exact match accuracy underestimates the model's true perceptual fidelity; evaluating at the broader regional level increases the specialist's accuracy to $0.91$, compared to $0.61$ for the frozen backbone.

\begin{table}[t]
\centering
\caption{The specialist mitigates output mode collapse without altering early perception. 
Accuracy of the
emitted textual attribute, frozen backbone vs.\ its specialist, on clips with ground-truth labels (word stress is scored
as the top stressed word against Whistress~\citep{whistress}).}
\label{tab:elicit_output}
\begin{tabular}{lcc}
\toprule
Attribute (metric) & Frozen backbone & + specialist \\
\midrule
Word stress (top-word accuracy) & 0.60 & 0.81 \\
Accent (exact, 13 classes)      & 0.37 & 0.62 \\
Accent (region)                 & 0.61 & 0.91 \\
Emotion (5 classes)             & 0.15 & 0.59 \\
\bottomrule
\end{tabular}
\end{table}

\section{IIO tasks}
\label{app:singlepass}

The unified student model is trained using the extension of our proposed Instruction-Input-Output (IIO) format. 
During training, each example is formatted as a single completion task, sampled according to fixed probabilities. 
The mixture comprises per-attribute tasks which describe isolated paralinguistic features and one \texttt{comprehensive} task designed to output the entire structured description at once. 
The comparative attribute specialists always receive the speaker turns and transcript as teacher-forced input.
Conversely, the \texttt{caption} and \texttt{comprehensive} tasks take no text input and decode directly from the raw audio (Table~\ref{tab:iio-mixture}). 
To establish single-pass capabilities, 
we append the \texttt{comprehensive} task (empty \texttt{INPUT}) at a weight of $1.0$ and reduced the \texttt{caption} task weight to $0.5$. 
The comprehensive task adds single-pass capability while retaining the cascaded multi-pass inference regime.

\begin{table}[t]\centering
\caption{\textbf{IIO tasks during training.} 
Each training sample is formulated as a single completion task, 
with weights denoting their respective sampling probabilities. 
The comparative speech specialists are conditioned on the speaker turns and transcript as teacher-forced inputs.}
\label{tab:iio-mixture}
\begin{tabular}{llcl}
\toprule
Task & Input & Weight & Output \\
\midrule
comprehensive & none (audio) & 1.0 & full description (all sections) \\
caption       & none (audio) & 0.5 & overview + events with transcript \\
stress        & transcript turns & 0.5 & per-word stress and intonation \\
emotion       & transcript turns & 0.5 & per-turn emotion \\
voice profile & transcript turns & 0.3 & per-speaker gender, age, accent \\
acoustics         & transcript turns & 0.3 & per-segment pitch and loudness \\
\bottomrule
\end{tabular}
\end{table}

At inference, the dynamic IIO surface allows the model to operate in two distinct regimes. 
The multi-pass approach executes a cascaded sequence of five decodes. First, the \texttt{caption} task processes the raw audio to generate the scene overview and the timestamped events list, including the verbatim transcript. 
The predicted speaker turns from this initial pass are subsequently fed back as the \texttt{INPUT} conditioning for the four specialist decodes (\texttt{stress}, \texttt{emotion}, \texttt{voice profile}, and \texttt{acoustics}). 
This \swap{composed} teacher-forcing ensures all fine-grained paralinguistic descriptors remain structurally aligned to the same master transcript.
Conversely, the single-pass approach utilizes only the \texttt{comprehensive} task. 
By reading the raw audio with an empty \texttt{INPUT} IIO, it emits the complete structured description, 
spanning all non-speech events as well as speech attributes, 
in a single, continuous output block. Both inference approaches yield the identical structured schema.

\section{Segment- and pause-level localization}
\label{app:localization}

\modelname{}'s training schema records a turn broken by a pause as a comma-separated list of active spans 
(for example, \texttt{<speech>...</speech> from 1.4s to 3.8s, 5.4s to 6.7s}).
We test whether this supervision produces accurate segment- and pause-level localization on real conversational speech, and 
how it compares to dedicated aligners. 

Specifically, we evaluate through the \texttt{caption} task (Appendix~\ref{app:singlepass})
wherein we mark segment boundaries in the reference transcript with the delimiter (\texttt{\textbackslash\textbackslash}), and 
recover each segment's predicted timestamps.
A segment is a run of speech bounded by a pause, whether a change of speaker or the same speaker pausing mid-utterance.

We use the Buckeye corpus of conversational American English~\cite{buckeye} (eight development speakers, 50 recordings).
We define a segment as a run of target-speaker words bounded by a silence of at least $0.1$\,s or an interviewer turn (provided in the dataset), and 
pack consecutive segments into windows of at most $20$\,s 
yielding $1{,}349$ windows and $4{,}947$ segments.

\begin{figure}[t]\centering
\includegraphics[width=\linewidth]{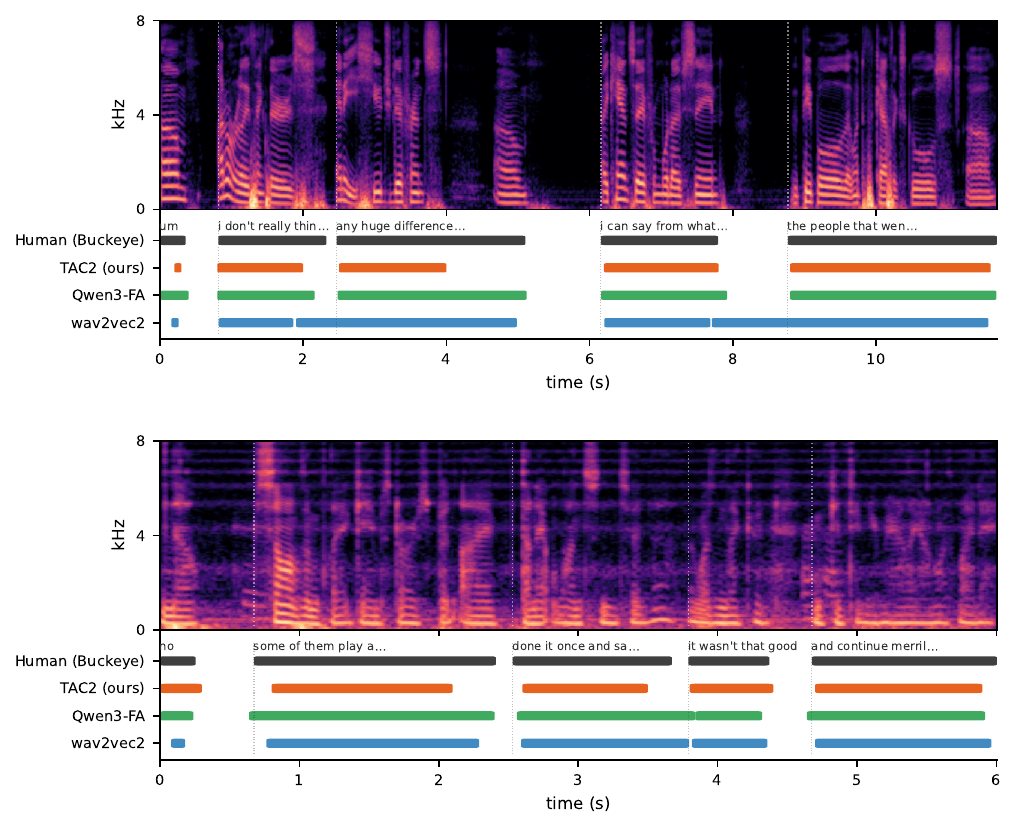}
\caption{
\textbf{Segment localization on conversational speech.} 
The segments from the human annotation, \modelname{}, the Qwen3-FA aligner, and the wav2vec2-based aligner share the time axis. \modelname{}'s segment onsets track the human pause boundaries closely, while being competitive with the dedicated aligners.
(Dotted lines mark the human segment onsets.)
}
\label{fig:seg_localization}
\end{figure}

For reference we run two specialized alignment models on the same windows, the Qwen3 aligner~\cite{Qwen3-ASR} and a wav2vec2-based aligner~\cite{torchaudio21}.
\modelname{} localizes segment onsets to a $49$\,ms median boundary error and 
attains the highest boundary agreement at a $200$\,ms tolerance ($0.93$), the most consistent of the three systems and 
competitive with models built solely for alignment (Table~\ref{tab:seg_localization}, Figure~\ref{fig:seg_localization}).
Importantly, \modelname{} itself proposes the segmentation, 
recovering the correct segment count $84\%$ of the time, 
whereas the aligners require this as input.

\begin{table}[t]
\centering
\footnotesize
\setlength{\tabcolsep}{6pt}
\renewcommand{\arraystretch}{1.15}
\caption{
\textbf{Segment- and pause-level localization on conversational speech.}  Median onset error (ms, $\downarrow$) and 
boundary F1 at 100/200\,ms ($\uparrow$).
\emph{Count}: fraction of windows where \modelname{} proposes the correct segment count.
}
\label{tab:seg_localization}
\begin{tabular}{lcccc}
\toprule
\textbf{System} & \textbf{Onset (ms)} & \textbf{F1@100} & \textbf{F1@200} & \textbf{Count} \\
\midrule
Qwen3-FA aligner~\cite{Qwen3-ASR}                 & \textbf{23} & \textbf{0.784} & 0.823 & n/a \\
wav2vec2 aligner~\cite{torchaudio21}  & 72 & 0.608 & 0.752 & n/a \\
\modelname{} (ours)                          & 49 & 0.774 & \textbf{0.929} & 84\% \\
\bottomrule
\end{tabular}
\end{table}

\input{sec/appendix_LSA}

\section{Extending the describer to audio-visual reasoning}
\label{app:tac2v}
The decoupled describe-then-reason paradigm is not limited to audio understanding but can readily be extended to serve as the acoustic foundation for an audio-visual describer. 
Following the TAC-V framework introduced by \citet{tac}, we generate an integrated audio-visual caption by pairing our audio describer with an off-the-shelf visual grounder (Qwen3-VL-32B). 
Specifically, the audio describer outputs timestamped acoustic events and speech, while a visual shot list delineates scene boundaries. 
The visual grounder then links acoustic events to their on-screen sources, 
leveraging visual context to resolve acoustic ambiguities. 
Finally, a frozen text reasoner answers downstream queries utilizing this unified multimodal caption alone.

We evaluate on four audio-visual reasoning benchmarks:
Daily-Omni~\citep{dailyomni},
WorldSense~\citep{worldsense},  
Video-Holmes~\citep{videoholmes}
and AVHBench~\citep{avhbench}.
Table~\ref{tab:tac2v} replicates the comparative setup of \citet{tac}, 
contrasting native multimodal large language models (which process audio and video directly) against the cascaded describe-then-reason approach. 
The \emph{VLM + Qwen3} ablation establishes a baseline using visual-only captions devoid of acoustic descriptions.
Substituting
\studentname{} improves the cascade on all three reasoning benchmarks ($+1.9$, $+0.4$, $+5.9$ over TAC-V) and beats every
native model on Daily-Omni and Video-Holmes.
On AVHBench, \studentname{}-V leads or matches TAC-V on the three \emph{discriminative} subtasks 
(audio-driven video hallucination AVH, 
video-driven audio hallucination VAH, and 
audio-visual matching AVM) but trails on the
\emph{generative} captioning subtask (AVC).
AVC evaluates a generated caption against a \emph{single} human reference using METEOR.
Interestingly, we find that even a faithful paraphrase of the reference itself attains only ${\sim}35$ METEOR. 
It is evident that wording differences alone therefore forfeit roughly
two-thirds of the achievable score, 
compressing the absolute range for every system.
We also noticed that the per-clip METEOR is high exactly when our caption's opening subject and word choice happens to coincide with
the reference and low otherwise.
We believe the metric tracks \emph{phrasing agreement}, 
not caption correctness.
We form \studentname{}-V by substituting \studentname{} for TAC's audio describer while keeping the same visual
grounder and the reasoner, so the improvement is only attributable to our generated audio description.

\begin{table}[t]
\centering
\caption{\textbf{Audio-visual reasoning performance.} 
Our \studentname{}-V pipeline substitutes \studentname{} for the prior TAC audio describer, keeping describe-then-reason pipeline constant.
Daily-Omni, World-Sense, and Video-Holmes report multiple-choice accuracy. 
AVHBench reports accuracy for AVH/VAH/AVM and 
METEOR for the AVC subtask. Best per row in bold.}
\label{tab:tac2v}
\resizebox{\linewidth}{!}{
\begin{tabular}{ll c | c | cc | cc}
        \toprule
        & \multicolumn{2}{c|}{\textbf{Native MLLM}} & \multicolumn{5}{c}{\textbf{Describe-then-reason}} \\
        \cmidrule(lr){2-3} \cmidrule(lr){4-8}
        \textbf{Benchmark} & Model & Score & \shortstack{VLM \\ + Qwen3} & \shortstack{TAC-V \\ + Qwen3} & \shortstack{TAC-V \\ + Gemini3}
        & \shortstack{\studentname{}-V \\ + Qwen3} & \shortstack{\studentname{}-V \\ + Gemini3} \\
        \midrule
        \multirow{3}{*}{Daily-Omni} & Qwen3-Omni & 76.2 & \multirow{3}{*}{51.5} & \multirow{3}{*}{72.9} & \multirow{3}{*}{77.9}
        & \multirow{3}{*}{77.2} & \multirow{3}{*}{\textbf{79.8}} \\
        & Gemini 2.5 Flash & 72.7 \\
        & OmniVinci & 66.5 \\
        \midrule
        \multirow{2}{*}{World-Sense} & Gemini 2.5 Pro & \textbf{65.1} & \multirow{2}{*}{37.4} & \multirow{2}{*}{45.7} & \multirow{2}{*}{58.6}
        & \multirow{2}{*}{55.8} & \multirow{2}{*}{59.0} \\
        & OmniVinci & 48.2 \\
        \midrule
        Video-Holmes & Qwen3-Omni & 57.3 & 45.6 & 47.7 & 59.2 & 58.4 & \textbf{65.1}\\
        \midrule
        AVHBench (AVH) & PandaGPT & 58.5 & 70.8 & 79.8 & 81.7 & 82.0 & \textbf{82.2} \\
        AVHBench (VAH) & PandaGPT & 61.3 & 51.8 & 76.1 & 76.6 & \textbf{76.9} & 76.4 \\
        AVHBench (AVM) & OneLLM & 60.1 & 50.5 & 56.7 & 61.6 & 72.6 & \textbf{77.5} \\
        AVHBench (AVC) & Video-LLaMa & 14.0 & 12.9 & \textbf{22.6} & 20.6 & 14.1 & 13.8 \\
        \bottomrule
        \end{tabular}
}
\end{table}

\section{Text to audio model and temporal grounding setup}
\label{tta_model}

\paragraph{Generator.} 
Derived from \citep{jin2025dtop}, our text-to-audio (T2A) architecture is a latent diffusion model that synthesizes 48~kHz audio conditioned on text captions. 
The base model is trained exclusively on structural \modelname{} captions,
which consist of a scene overview and a timestamped event list devoid of speaker characteristics. 
A subsequent fine-tuning phase incorporates enriched \studentname{} captions (include speaker demographics and word-level emphasis tags), that enable explicit control over speaker attributes while maintaining the temporal fidelity (see Section~\ref{sec:exp_generative}). 
To evaluate the base T2A model, we deterministically ablate speaker metadata and emphasis tags from the enriched captions, 
retaining only the scene overview and timestamped events.

\paragraph{\studentname{}-as-judge.}
Beyond timing, we test whether the full schema survives synthesis with a describe--generate--re-describe approach.
We re-caption the generated audio with \studentname{} and match the re-caption to the input caption field by field.
We report accuracy for speaker gender, age, and accent and for emotion, recall for all events and for speech events, and transcript word error rate.
A missing or mismatched attribute counts as an error, so every field is scored over the full clip set.
This tests whether the commanded speaker attributes, events, and spoken content are recoverable from the generated audio, complementing the temporal and perceptual metrics below.

\paragraph{Temporal Grounding Evaluation.} 
We construct a diverse evaluation suite of 300 held-out English clips encompassing six accents 
(United States, United Kingdom, Nigeria, India, Australia, and the Philippines). 
Each clip contains between three and ten distinct acoustic events, yielding 1,471 total events. 
To measure temporal fidelity without treating the generated text caption as a reliable ground truth, 
we utilize the original recording as a zero-distance reference. 
We employ FLAM~\citep{flam} and PE-A-Frame~\citep{peaframe}, 
two frame-level audio-text models, 
to compute semantic similarity scores for every short-time frame against a given text query, 
extracting the peak activation frame as the predicted temporal location. 
For each event, we report the absolute temporal offset (in seconds) between the localizer's peak in the generated audio and its corresponding peak in the original recording. 
This error is averaged exclusively over the events successfully localized in the original recording to ensure a reliable reference.

The peak-offset metrics localize each event against a text query.
We additionally report a fully text-free measure: 
using PE-A-Frame's audio tower alone, 
we embed the original and generated audio into one $1024$-d vector per $40$\,ms frame and report their mean per-frame cosine distance averaged over clips.
The distance to original audio is zero, as reference.
We believe this is a recall-heavy, global reconstruction measure.
It ranks the systems identically to the peak-offset metrics and tracks the Gemini-3-Pro judge, 
supporting the view that the \studentname{} caption is a compact latent expressed in text from which the scene can be re-rendered.

\paragraph{Baselines.} 
We benchmark against 
MiDashengLM-Gen~\citep{midashenglmgen}, a recent text-to-audio scene generator and 
PicoAudio2~\citep{picoaudio2}, a model explicitly designed for timestamp control, evaluating it both with and without event timestamps to provide a clean intra-model ablation. 
Note that PicoAudio2 does not model speech. 
Finally, we evaluate Seed Audio~1.0~\citep{seedaudio}, 
a highly capable commercial text-to-audio model prompted with the identical caption formatted as a scene overview followed by timestamped event lines. 

\section{Prompts}
\label{sec:prompts}
For MMSU, MMAR, and MMAU, for the audio-visual benchmarks, and for the open-weight reasoners in our robustness study (Section~\ref{sec:exp_singlepass}), we use the describe-then-reason prompts of \citet{tac}. 
Below we provide the two prompts specific to this work: the describe-then-reason wrapper for MSU-Bench, and the audio-native judge used to evaluate text-to-audio synthesis.

\subsection{MSU-Bench reasoner}
We score MSU-Bench with its official evaluation harness~\citep{msubench}, 
using its multiple-choice prompt (the question, the answer options, and a fixed output-format instruction) verbatim. 
We replace the audio with \studentname{}'s text description of the segment(s) referenced by each question, prepending it as shown in Figure~\ref{fig:msubench_prompt}.
The reasoner has access to only the caption generated by our models.

\begin{figure}[ht]
\centering
\begin{tcolorbox}[colback=gray!5!white, colframe=black!75!black, title=\textbf{Describe-then-reason prompt for MSU-Bench}]
\small
\fontfamily{pcr}\selectfont
Audio description:\\
\{description\}\\[6pt]
\{question\}\\[6pt]
Options:\\
\{options\}\\[6pt]
{}[Output format]\\
On the very last line of your reply, output only: Final answer: A/B/C/D.
\end{tcolorbox}
\caption{The MSU-Bench describe-then-reason prompt. 
\texttt{\{description\}} is \studentname{}'s text description of the referenced segment(s).
The output-format instruction is the official MSU-Bench harness prompt~\citep{msubench} used verbatim.}
\label{fig:msubench_prompt}
\end{figure}

\subsection{Text-to-audio perceptual judge (Gemini-3-Pro)}
To conduct holistic perceptual evaluations, we employ Gemini-3-Pro as an audio-native judge. 
The model is queried greedily with one call per (system, clip) pair across the evaluation set. 
The judge evaluates systems across two rubrics. 
\emph{vs.\ real} passes both the original recording and the generated candidate, prompting the model to rate their acoustic and thematic resemblance on a 1-5 scale. 
\emph{vs.\ caption} passes the enriched \studentname{} text caption alongside the generated audio, prompting the model to evaluate how faithfully the candidate synthesizes the requested description. 
Since every system is judged against the \emph{same} \studentname{} caption, this metric provides a direct, level-field cross-model comparison.

\begin{figure}[ht]
\centering
\begin{tcolorbox}[colback=gray!5!white, colframe=black!75!black, title=\textbf{Text-to-audio judge: \emph{vs.\ real} recording}]
\footnotesize
\fontfamily{pcr}\selectfont
You are an expert audio evaluator. You are given two short audio clips: REFERENCE, an original real-world recording, and CANDIDATE, audio from a text-to-audio model asked to reproduce the same scene. Rate how closely the CANDIDATE resembles the REFERENCE on an integer scale 1-5: 5 = near-identical (same distinct sound events at essentially the same times, matching speaker/voice characteristics, scene, and ambience); 4 = strong (most events present and roughly aligned; minor differences); 3 = moderate (general scene recognizable but several events missing, mistimed, or altered); 2 = weak (loose thematic connection only); 1 = no resemblance. Weigh, in order: (a) the same events are present, (b) their timing and order, (c) speaker/voice characteristics for any speech (gender, age, accent, delivery), (d) overall scene, ambience, and recording quality. Reason briefly, then end with a JSON object on its own final line: \{"rating": <1-5>\}.
\end{tcolorbox}
\caption{Gemini-3-Pro audio-native judge prompt for the \emph{vs.\ real} rubric.}
\label{fig:judge_vs_real}
\end{figure}

\begin{figure}[ht]
\centering
\begin{tcolorbox}[colback=gray!5!white, colframe=black!75!black, title=\textbf{Text-to-audio judge: \emph{vs.\ caption}}]
\footnotesize
\fontfamily{pcr}\selectfont
You are an expert audio evaluator. You are given a DESCRIPTION, a structured text description of an intended audio clip (it may contain \# Overview / \# Events sections, per-event speaker attributes, timestamps of the form `from Xs to Ys', a transcript inside <speech>...</speech>, and inline emphasis markup <e>word</e> / <E>word</E>), and a CANDIDATE, audio produced by a text-to-audio model conditioned on this description. Rate how faithfully the CANDIDATE realizes the DESCRIPTION on an integer scale 1-5: 5 = faithful (all described events at the described times and order; for speech, the words, described speaker attributes, and any marked emphasis all match; scene and ambience match); 4 = strong (most described content present and roughly aligned); 3 = moderate (scene recognizable but several described events, attributes, or timings missing or wrong); 2 = weak (loose connection only); 1 = unrelated. Weigh, in order: (a) the described events are present, (b) their timing and order, (c) for speech, the spoken content and described speaker attributes and marked emphasis, (d) overall scene, ambience, and quality. Judge only against the DESCRIPTION; do not penalize plausible details it leaves unspecified. Reason briefly, then end with a JSON object on its own final line: \{"rating": <1-5>\}.
\end{tcolorbox}
\caption{Gemini-3-Pro audio-native judge prompt for the \emph{vs.\ caption} rubric. Since every system is judged against the same \studentname{} caption, this yields a level-field cross-model comparison.}
\label{fig:judge_vs_caption}
\end{figure}

%% file: sec/appendix_LSA.tex
\begin{swapblock}
\section{Long-form audio understanding}
\label{app:longaudio}

We test whether \modelname{}, trained on clips of at most 30\,s, extends to recordings of up to 2.7\,h on LongAudioSpan~\citep{longaudiospan}.
It contains 360 English and Chinese recordings in three duration tiers, with questions scored in three modes:
multiple-choice \emph{Accuracy},
open-ended \emph{Rubric}, where an LLM judge checks both content and timing, and
\emph{Chain}, three linked questions about an edit inserted into the audio.
We describe each recording in chunks of at most 30\,s and join the descriptions on one timeline.
A frozen text-only GPT-5.4 answers from this description alone, using the benchmark's prompts and official scorers.

\begin{table*}[t]
\centering
\setlength{\tabcolsep}{3pt}
\renewcommand{\arraystretch}{1.15}
\def\tr#1{\textcolor{gray}{#1}}
\caption{
\textbf{Long-form audio understanding on LongAudioSpan~\citep{longaudiospan}.}
Accuracy, Rubric and Chain over duration tiers S (10 to 30\,min), M (30 to 60\,min) and L (over 60\,min); Avg is the mean over tiers.
Baseline rows are from \citet{longaudiospan}; \textcolor{gray}{gray} cells use audio truncated to the model's input Limit (minutes).
For Chain, the caption baseline sees the edits written into its caption.
Our row pairs \modelname{} (Size) with a frozen text-only GPT-5.4 reasoner.
Best in \textbf{bold}, second \underline{underlined}, among audio models and ours.
}
\resizebox{\textwidth}{!}{%
\begin{tabular}{l c c c cccc cccc cccc}
\toprule
& & & & \multicolumn{4}{c}{\textbf{Accuracy (\%)}} & \multicolumn{4}{c}{\textbf{Rubric (\%)}} & \multicolumn{4}{c}{\textbf{Chain (\%)}} \\
\cmidrule(lr){5-8} \cmidrule(lr){9-12} \cmidrule(lr){13-16}
\textbf{Model} & \textbf{Size} & \textbf{Think} & \textbf{Limit} & S & M & L & Avg & S & M & L & Avg & S & M & L & Avg \\
\midrule
\multicolumn{16}{c}{\textit{Text-only Baselines}} \\
\midrule
GPT-5.4 (question-only) & -- & \cmark & -- & 41.1 & 41.1 & 40.6 & 40.9 & -- & -- & -- & -- & 8.9 & 11.9 & 10.3 & 10.4 \\
GPT-5.4 (transcript) & -- & \cmark & -- & 64.8 & 60.7 & 53.4 & 59.7 & 41.0 & 41.5 & 37.2 & 39.9 & 9.4 & 14.2 & 14.2 & 12.6 \\
GPT-5.4 (caption) & -- & \cmark & -- & 83.2 & 80.0 & 77.3 & 80.2 & 59.1 & 57.4 & 53.0 & 56.5 & 66.7 & 64.4 & 53.1 & 61.4 \\
\midrule
\multicolumn{16}{c}{\textit{Open-source Models}} \\
\midrule
Phi-4-Multimodal & 5.6B & \xmark & 30m & 33.7 & \tr{34.6} & \tr{34.2} & 34.2 & 18.4 & \tr{13.6} & \tr{10.1} & 14.0 & 12.9 & \tr{7.2} & \tr{8.8} & 9.6 \\
\multirow{2}{*}{AudioFlamingoNext} & \multirow{2}{*}{8B} & \xmark & 30m & 46.1 & \tr{43.6} & \tr{37.2} & 42.3 & 16.1 & \tr{12.2} & \tr{13.7} & 14.0 & 5.3 & \tr{5.6} & \tr{6.1} & 5.7 \\
 & & \cmark & 30m & 43.3 & \tr{38.9} & \tr{35.0} & 39.1 & 26.9 & \tr{20.3} & \tr{16.3} & 21.2 & 7.5 & \tr{11.1} & \tr{13.1} & 10.6 \\
Voxtral-Mini & 3B & \xmark & 40m & 45.3 & \tr{35.8} & \tr{37.2} & 39.4 & 25.6 & \tr{21.8} & \tr{13.6} & 20.3 & 5.6 & \tr{9.4} & \tr{18.3} & 11.1 \\
Voxtral-Small & 24B & \xmark & 40m & 42.5 & \tr{40.0} & \tr{31.9} & 38.1 & 28.9 & \tr{24.2} & \tr{20.7} & 24.6 & 12.5 & \tr{4.7} & \tr{4.7} & 7.3 \\
\multirow{2}{*}{MOSS-Audio} & \multirow{2}{*}{8B} & \xmark & 50m & 45.0 & \tr{43.1} & \tr{35.6} & 41.2 & 23.7 & \tr{17.7} & \tr{10.7} & 17.4 & 8.3 & \tr{11.4} & \tr{9.2} & 9.6 \\
 & & \cmark & 50m & 52.2 & \tr{52.8} & \tr{38.6} & 47.9 & 30.1 & \tr{15.4} & \tr{11.6} & 19.0 & 13.9 & \tr{15.3} & \tr{11.4} & 13.5 \\
Baichuan-Omni-1.5 & 11B & \xmark & 80m & 46.9 & 46.9 & \tr{36.4} & 43.4 & 19.0 & 12.5 & \tr{9.7} & 13.7 & 11.9 & 14.4 & \tr{14.7} & 13.7 \\
\multirow{2}{*}{Qwen3-Omni} & \multirow{2}{*}{30B (A3B)} & \xmark & 80m & 53.9 & 54.4 & \tr{46.2} & 51.5 & 31.3 & 24.3 & \tr{20.6} & 25.4 & 20.5 & 19.2 & \tr{13.2} & 17.6 \\
 & & \cmark & 80m & 56.9 & 57.5 & \tr{49.2} & 54.5 & 38.4 & 29.2 & \tr{22.8} & 30.1 & 7.8 & 10.6 & \tr{15.3} & 11.2 \\
\midrule
\multicolumn{16}{c}{\textit{Closed-source Models}} \\
\midrule
Gemini-2.5-Pro & -- & \cmark & -- & 67.5 & 64.2 & 54.7 & 62.1 & 46.9 & 40.7 & 36.4 & 41.3 & 24.2 & 25.3 & 21.9 & 23.8 \\
Gemini-3-Flash & -- & \xmark & -- & 65.6 & 69.2 & 64.2 & 66.3 & 54.9 & \underline{55.0} & \underline{53.8} & \underline{54.6} & 16.7 & 20.6 & 11.9 & 16.4 \\
Gemini-3.1-Pro & -- & \cmark & -- & 76.7 & \underline{74.4} & 65.8 & 72.3 & 56.2 & 48.8 & 43.3 & 49.4 & 17.5 & 25.3 & 20.0 & 20.9 \\
Qwen3.5-Omni-Flash & -- & \xmark & -- & 73.9 & 71.4 & 58.9 & 68.1 & 50.3 & 40.9 & 35.9 & 42.4 & 23.1 & 19.2 & 18.3 & 20.2 \\
Qwen3.5-Omni-Plus & -- & \xmark & -- & \underline{79.2} & \underline{74.4} & \underline{70.3} & \underline{74.6} & \underline{56.9} & 48.6 & 47.0 & 50.8 & \underline{30.8} & \underline{33.1} & \underline{23.1} & \underline{29.0} \\
\midrule
\multicolumn{16}{c}{\textit{Describe-then-Reason (ours)}} \\
\midrule
\modelname{} $\rightarrow$ GPT-5.4 (text-only) & 30B (A3B) & \cmark & -- & \textbf{83.1} & \textbf{79.4} & \textbf{74.4} & \textbf{79.0} & \textbf{76.3} & \textbf{71.6} & \textbf{68.7} & \textbf{72.2} & \textbf{32.8} & \textbf{35.0} & \textbf{31.4} & \textbf{33.1} \\
\bottomrule
\end{tabular}%
}
\label{tab:longaudiospan}
\end{table*}

\modelname{} outperforms every audio model in all three modes and tiers (Table~\ref{tab:longaudiospan}),
and on Accuracy nearly matches the dense caption the questions were written from (79.0 vs.\ 80.2).
The largest gain is on Rubric, which asks \emph{when} events happen.
For closed-source audio models, the share of correctly placed timestamps falls from 42\% on the shortest tier to 23\% on the longest~\citep{longaudiospan},
whereas reading the \modelname{} timeline, the reasoner places 78\% and 70\% correctly.
On Chain, understanding answers are 17 points more accurate when the edit is located than when it is missed,
more than for any closed-source audio model (at most 13 points), so the credit comes from grounding rather than guessing.
\end{swapblock}